\documentclass[10pt]{article}
\usepackage[margin=3cm]{geometry}

\usepackage{natbib}
\usepackage{eqnarray,amsmath}
\usepackage{mathtools}
\usepackage{amssymb}
\usepackage{amsthm}
\usepackage{mathtools}
\usepackage{graphicx}
\usepackage{epstopdf}
\usepackage[linesnumbered]{algorithm2e}
\usepackage{bm}
\usepackage{multirow} 
\usepackage{color}
\usepackage[toc,page]{appendix}
\usepackage{url}
\usepackage{fixltx2e}
\usepackage{hyperref}

\usepackage{subcaption}

\usepackage{todonotes}

\usepackage{authblk}

\usepackage{bbm} 

\usepackage{pifont}

\DeclareMathOperator*{\argmin}{arg\,min}

\usepackage{lscape}

\title{Regularized Zero-Variance Control Variates}
\author[$\dag,\dag\dag$]{L. F. South}
\author[$\dag\dag\dag$]{C. J. Oates}
\author[$\dag\dag\dag\dag$]{A. Mira}
\author[$\dag$]{C. Drovandi}
\affil[$\dag$]{School of Mathematical Sciences, Queensland University of Technology, Australia\\
ARC Centre of Excellence for Mathematical \& Statistical Frontiers (ACEMS)}
\affil[$\dag\dag$]{Department of Mathematics and Statistics, Lancaster University, United Kingdom}
\affil[$\dag\dag\dag$]{School of Mathematics, Statistics and Physics, Newcastle University, UK\\
Alan Turing Institute, UK}
\affil[$\dag\dag\dag\dag$]{Faculty of Economics, Universit\`{a} della Svizzera italiana, Switzerland\\
University of Insubria, Italy}

\begin{document}
\newcommand{\vect}[1]{\boldsymbol{#1}}
\newcommand{\vtheta}{\boldsymbol{\theta}}
\newcommand{\zv}[1]{$\text{ZV}_{#1}$}
\newcommand{\zvl}[1]{$l\text{-ZV}_{#1}$}
\newcommand{\zvr}[1]{$r\text{-ZV}_{#1}$}
\newcommand{\zvs}[2]{$\text{sub}_{#2}\text{-ZV}_{#1}$}
\newcommand{\zvsl}[2]{$\text{sub}_{#2}\text{-}l\text{-ZV}_{#1}$}
\newcommand{\zvsr}[2]{$\text{sub}_{#2}\text{-}r\text{-ZV}_{#1}$}

\newcommand{\ntimes}{{\mkern-2mu\times\mkern-2mu}}

\maketitle

\setlength{\parindent}{0pc}
\setlength{\parskip}{1ex}

\begin{abstract}

Zero-variance control variates (ZV-CV) are a post-processing method to reduce the variance of Monte Carlo estimators of expectations using the derivatives of the log target. Once the derivatives are available, the only additional computational effort lies in solving a linear regression problem. Significant variance reductions have been achieved with this method in low dimensional examples, but the number of covariates in the regression rapidly increases with the dimension of the target. In this paper, we present compelling empirical evidence that the use of penalized regression techniques in the selection of high-dimensional control variates provides performance gains over the classical least squares method. Another type of regularization based on using subsets of derivatives, or \textit{a priori} regularization as we refer to it in this paper, is also proposed to reduce computational and storage requirements. Several examples showing the utility and limitations of regularized ZV-CV for Bayesian inference are given. The methods proposed in this paper are accessible through the R package \textsf{ZVCV}.

\end{abstract}


\section{Introduction}
\label{sec:intro}

Our focus in this paper is on calculating the expectation of a square integrable function $\varphi(\vtheta)$ with respect to a distribution with (Lebesgue) density $p(\vtheta)$, $\vtheta \in \Theta \subseteq \mathbb{R}^d$. Given independent and identically distributed (iid) samples $\{\vtheta_i\}_{i=1}^{N} \stackrel{\text{iid}}{\sim} p(\vtheta)$, the standard Monte Carlo estimator,
\begin{equation}\label{eqn:MonteCarlo}
\widehat{\mathbb{E}_{p}[\varphi(\vtheta)]} =\frac{1}{N}\sum_{i=1}^N \varphi(\vtheta_i),
\end{equation}
is an unbiased estimator of $\mathbb{E}_{p}[\varphi(\vtheta)] = \int_\Theta \varphi(\vtheta) p(\vtheta) \mathrm{d} \vtheta$ and its variance is $\mathcal{O}(1/N)$. Reducing the variance of this estimator by increasing $N$ is often infeasible due to the cost of sampling from $p(\vtheta)$ and potentially the cost of evaluating $\varphi(\vtheta)$. If the samples are not iid then the functional form of the estimator is the same and the methods described in this work still apply.

Recent control variate methods have focused on reducing the variance of \eqref{eqn:MonteCarlo} using the derivatives of the log target, $\nabla_{\vtheta} \log{p(\vtheta)}$, or some unbiased estimator of this quantity. Zero-variance control variates (ZV-CV) \citep{Assaraf1999,Mira2013} and control functionals (CF) \citep{Oates2017a} are two such methods. ZV-CV amounts to solving a linear regression problem and CF is a non-parametric alternative. These methods can be used as post-processing procedures after $N$ samples, not necessarily iid, from $p$ have been produced along with evaluations of $\nabla_{\vtheta}\log{p(\vtheta)}$ and $\varphi(\vtheta)$ for each of the samples. Often $\nabla_{\vtheta}\log{p(\vtheta)}$ is already available because derivative-based methods like Metropolis adjusted Langevin algorithm (MALA) \citep{Roberts2002,Girolami2011} or Hamiltonian Monte Carlo (HMC) \citep{Duane1987,Girolami2011} have been used in the sampling algorithm.

The parametric approximation in ZV-CV is based on a polynomial in $\vtheta$, so the number of coefficients to estimate rapidly increases both with the polynomial order and with $d$. As a result of restricting to a low polynomial order, ZV-CV tends to offer less substantial improvements than CF for challenging low-dimensional applications. This is not surprising given the good statistical properties of CF which have been described in \citet{Oates2017,Barp2018}. However, CF has an $\mathcal{O}(N^3)$ computational cost, compared to ZV-CV with has computational cost of $\mathcal{O}(N)$, and it also suffers from the curse of dimensionality with respect to $d$ due to the use of non-parametric methods. Some results in \citet{Oates2017a}, shown mainly in the appendices, suggest that the performance of CF compared to ZV-CV may deteriorate in higher dimensions.

One aim of this work is to develop derivative-based control variate methods which are inexpensive, effective and capable of handling higher dimensions than existing derivative-based methods. The novel methods that we introduce are referred to as \textit{regularized} ZV-CV and they are based on two types of regularization: penalization methods for linear regression and what we refer to as \textit{a priori} regularization. Penalized ZV-CV allows higher order polynomials to be used than could be considered with ordinary least squares. This method is motivated by showing that $\mathcal{L}_2$ penalized ZV-CV is equivalent to CF with a second-order differential operator and finite-dimensional polynomial kernel. \textit{A priori} ZV-CV is most beneficial when $N < d$. Empirical results in Section \ref{sec:Examples} suggest that significant variance reductions can be achieved with \textit{a priori} ZV-CV when $N<d$ or with penalized ZV-CV when the polynomial order is pushed beyond the limits of what standard ZV-CV can handle. We have developed an R package, \textsf{ZVCV} \citep{R_ZVCV}, which implements several derivative-based variance reduction techniques including standard ZV-CV, CF and regularized ZV-CV.

An important application area for ZV-CV and regularized ZV-CV is Bayesian inference, where Monte Carlo integration is commonly used. The use of ZV-CV and CF to improve posterior expectations based on samples from Markov chain Monte Carlo (MCMC) \citep{Metropolis1953} is well established (see e.g.\ \citet{Mira2013,Papamarkou2014,Oates2017a}). ZV-CV and CF have also been applied to the power posterior \citep{Friel2008} estimator of the normalising constant in an MCMC setting by \citet{Oates2016} and \citet{Oates2017a}, where they refer to this method as controlled thermodynamic integration (CTI). In this paper we go beyond existing literature and describe how regularized ZV-CV fits naturally into the context of sequential Monte Carlo (SMC) samplers \citep{DelMoral2006,Chopin2002}. In doing so, we provide a setting where adaptive methods can easily be applied to the CTI estimator. A novel reduced-variance normalising constant estimator using the standard SMC identity is also proposed. 

An introduction to derivative-based Monte Carlo variance reduction methods is provided in Section \ref{sec:Background}. The main methodological contributions in terms of developing regularized ZV-CV methods can be found in Section \ref{sec:Regularization}. Section \ref{sec:Examples} contains a simulation study comparing methods and estimators on the novel application to SMC. A final discussion of limitations and possible future work is given in Section \ref{sec:Discussion}.

\section{Control Variates based on Stein Operators}\label{sec:Background}

In this section, we recall previous work on control variate methods. The classical framework for control variates \citep{Ripley1987,Hammersley1964} is to determine an auxiliary function $\tilde{\varphi}(\vtheta) = \varphi(\vtheta) + h(\vtheta)$ such that $\mathbb{E}_{p}[\tilde{\varphi}(\vtheta)]=\mathbb{E}_{p}[\varphi(\vtheta)]$ and $\mathbb{V}_{p}[\tilde{\varphi}(\vtheta)]<\mathbb{V}_{p}[\varphi(\vtheta)]$, where $\mathbb{V}_{p}$ denotes the variance with respect to $p(\vtheta)$. Estimator \eqref{eqn:MonteCarlo} can then be replaced with the unbiased, reduced variance estimator,
\begin{equation}\label{eqn:MCwithCV}
\widehat{\mathbb{E}_{p}[\varphi(\vtheta)]} =\frac{1}{N}\sum_{i=1}^N \left[ \varphi(\vtheta_i) + h(\vtheta_i) \right].
\end{equation}

A control variate which has been considered in \citep{Assaraf1999,Mira2013,Barp2018} is
\begin{equation}\label{eqn:Stein}
\begin{aligned}
h_g(\vtheta) &= \mathcal{L}g(\vtheta) = \frac{\nabla_{\vtheta} \cdot (p(\vtheta) \nabla_{\vtheta}g(\vtheta))}{p(\vtheta)} \\
&= \Delta_{\vtheta}g(\vtheta) + \nabla_{\vtheta} g(\vtheta) \cdot \nabla_{\vtheta} \log{p(\vtheta)},
\end{aligned}
\end{equation}
where $\mathcal{L}$ is a second-order Langevin Stein operator \citep{Stein1972,Gorham2015} depending on $p$, $\Delta_{\vtheta}$ is the Laplacian operator represented in coordinates as $\sum_{j=1}^d \nabla_{\vtheta[j]}^2$ on $\mathbb{R}^d$, $\cdot$ is the dot product operator such that $\mathbf{a} \cdot \mathbf{b} = \mathbf{a}^{\top} \mathbf{b}$ and $g :\Theta \rightarrow \mathbb{R}$ is a twice continuously differentiable function to be specified.

Langevin Stein operators are helpful in generating control variates for two reasons. Firstly, they can be applied in Bayesian inference because they do not require the normalising constant of $p$. Furthermore, by definition a Stein operator $\mathcal{L}$ depending on $p(\vtheta)$ satisfies $\mathbb{E}_{p}[\mathcal{L} g(\vtheta)]=0$ for all functions $g(\vtheta)$ in a set called a Stein function class (see Section 2.2) and therefore $\mathbb{E}_{p}[\tilde{\varphi}(\vtheta)]=\mathbb{E}_{p}[\varphi(\vtheta)]$ under mild conditions. Typically it is also a requirement that Stein operators $\mathcal{L}$ fully characterize $p$ but this is not necessary for Stein-based control variates.

What remains is to choose $g$. The function $g$ for which $\tilde{\varphi}(\vtheta)$ is constant,  and thus zero variance is achieved, is generally intractable. In practice, $g$ is restricted to some function class $\mathcal{G}$ and is estimated based on samples targeting $p$.

\subsection{Choice of Function $g$}\label{ssec:var_red}
Variance reduction is effected through judicious choice of $g$. Once a function class $\mathcal{G}$ has been selected, the function $g \in \mathcal{G}$ is estimated by performing a regression task. As described in \citet{Barp2018}, a generalization of several existing regression methods for this problem is
\begin{equation}
(\hat{c},\hat{g}) \in \argmin_{\substack{c \in \mathbb{R} \\ g \in \mathcal{G}}} \frac{1}{N} \sum_{i=1}^N \left[ \varphi(\bm{\theta}_i) - c + \mathcal{L}g(\bm{\theta}_i)\right]^2 + \lambda\text{PEN}(g), \label{eqn:GenOpt}
\end{equation}
where $\text{PEN}(g)$ is a penalty function to be specified and $\lambda\geq 0$. This amounts to a penalized least squares approach to estimating $\varphi(\vtheta)$ using the functional form $c - \mathcal{L}g(\vtheta)$. This perspective on the optimization problem encompases ZV-CV, CF and neural control variates \citep{Zhu2018} as special cases. Further details on how ZV-CV and CF fit into this framework are given below. The main developments in this paper are based on considering alternative penalty functions.

Two recent contributions in control variates have optimization functions which do not fit into this framework, though the developments in penalty functions that are proposed in this paper could still be considered in these alternative frameworks. \citet{Belomestny2017} consider empirical variance minimization since minimising a square error objective function may not be optimal. \citet{Brosse2019} consider an alternative optimization problem which is motivated by minimising the asymptotic variance of a Langevin diffusion, which may be more suitable when samples have been obtained using MCMC with multivariate normal random walk or MALA proposals.

\subsubsection{Control Functionals}

CF \citep{Oates2017a,Barp2018} is based on choosing $\mathcal{G} \equiv \mathcal{H}$ where $\mathcal{H}$ is a user-specified Hilbert space of twice differentiable functionals on $\Theta$. The penalty term $\text{PEN}(g)$ considered in CF is $\text{PEN}(g) = \|g\|_{\mathcal{H}}^2$, where $\|\cdot\|_{\mathcal{H}}$ is the norm associated with the Hilbert space $\mathcal{H}$. The existence of a solution pair $(\hat{c},\hat{g}) \in \mathbb{R} \times \mathcal{H}$, together with an explicit algorithm for its computation, was obtained in that work under the assumption that the Hilbert space $\mathcal{H}$ admits a reproducing kernel (see \citet{Berlinet2011} for background). This method leads to estimators with super-root-$N$ convergence under conditions described in \citet{Oates2017} and \citet{Barp2018}. However, the cost associated with computation of $\hat{g}$ is $O(N^3)$, due to the need to invert a dense kernel matrix, and moreover this matrix is typically not well-conditioned. For applications that involve MCMC and SMC, typically $N$ will be at least $10^3$ and thus (in the absence of further approximations) the algorithm of  \citet{Oates2017a,Barp2018} can become impractical.

\subsubsection{Zero-Variance Control Variates}\label{ssec:ZVCV}

ZV-CV \citep{Assaraf1999,Mira2013} amounts to using $\mathcal{G}$ as the class of $Q$th order polynomial functions in $\vtheta$, and $\lambda = 0$. The polynomials $P(\vtheta)$ that we consider have total degree $Q\in \mathbb{Z}_{\geq 0}$, meaning that the maximum sum of exponents is $Q$ and the monomial basis is $\theta[1]^{\alpha_{1}}\cdots \theta[d]^{\alpha_{d}}$ where $\sum_{j=1}^d \alpha_j \leq Q$ and $\alpha \in \mathbb{Z}_{\geq 0}^d$. Substituting $g(\vtheta) = P(\vtheta) = \sum_{j=1}^J \beta_j P_j(\vtheta)$ into \eqref{eqn:Stein}, where $P_j(\vtheta)$ is the $j$th monomial in the polynomial and $\vect{\beta} \in \mathbb{R}^{J}$ is the vector of polynomial coefficients, gives
\begin{align*}
h_P(\vtheta) &= \mathcal{L}P(\vtheta)\\
&=\sum_{j=1}^J\beta_j\mathcal{L}P_j(\vtheta)\\
&= \vect{\beta}^\top\vect{x}(\vtheta).
\end{align*}
The $j$th zerovariate (covariate in the regression), $x_j(\vtheta) = \mathcal{L} P_j(\vtheta)$ is a term containing $\vtheta$ and $\nabla_{\vtheta} \log p(\vtheta)$. Its exact form is given in Appendix A of the Online Resources. For a $Q$th order polynomial when the dimension of $\vtheta$ is $d$, the constant $J={{d + Q}\choose{d}}-1$ is the number of regression parameters, excluding the intercept which is in the null space of the Stein operator.

The standard approach in the literature for choosing $\vect{\beta}$ is to perform ordinary least squares (OLS) \citep{Glasserman2003}. This is equivalent to choosing $\lambda = 0$ in \eqref{eqn:GenOpt}. The computational cost of ZV-CV is $\mathcal{O}(J^3 + NJ^2)$, which scales better with $N$ than CF which has computational cost $\mathcal{O}(N^3)$, where often $J \ll N$. Unlike in CF, regularization methods have not previously been used in connection with ZV-CV. 

Common practice is to default to $Q=2$ in ZV-CV. \citet{Mira2013} consider $Q=1$ to at most $Q=3$ and find that $Q=2$ is sufficient to achieve orders of magnitude variance reduction in most of their examples. \citet{Papamarkou2014} consider $Q\leq2$, pointing out that ``first and second degree polynomials suffice to attain considerable variance reduction.'' Low polynomial orders are also typically used in applications, for example \citet{Baker2017} use $Q=1$ and \citet{Oates2016} use $Q\leq2$. \citet{Oates2017a} compare CF with ZV-CV using $Q=2$ in most examples. 

It has previously been proposed to increase the number of control variates as the sample size increases (see e.g.\ \citet{Portier2018} and the appendices of \citet{Oates2017a}). This approach can be motivated by the Stone-Weierstrass theorem \citep{Stone1948}, which states that polynomial functions can be used to uniformly approximate, to an arbitrary level of precision, continuous functions on closed intervals. However, the increased number of coefficients in higher order polynomials may not be feasible or efficient to estimate with OLS. 

We demonstrate in Section \ref{sec:Examples} that the common practice of defaulting to $Q=2$ with OLS is often sub-optimal. Regularization approaches are proposed in Section \ref{sec:Regularization} to enable higher degree polynomials to be employed whilst avoiding over-fitting of the regression model used to estimate the  optimal coefficients of the polynomials.

\subsection{Unbiasedness}\label{ssec:Unbiasedness}

Suppose that $g(\vtheta)$ and $p(\vtheta)$ are sufficiently regular so that $\log p(\vtheta)$ has continuous first order derivatives and $g(\vtheta)$ has continuous first and second order derivatives. Also suppose that, if $g$ is to be estimated, then the samples used in estimating $g(\vtheta)$ are independent of those used in evaluating \eqref{eqn:MCwithCV}. If $\Theta \neq \mathbb{R}^d$, then we require that $\Theta$ is compact and has piecewise smooth boundary $\partial\Theta$. Under these conditions, estimator \eqref{eqn:MCwithCV} with samples $\{\vtheta_i\}_{i=1}^{N} \stackrel{\text{iid}}{\sim} p(\vtheta)$ is unbiased if
\begin{equation}\label{eqn:Boundary}
\oint_{\partial \Theta} p(\vtheta)\nabla_{\vtheta} g(\vtheta)\cdot \vect{n}(\vtheta) S(\mathrm{d}\vtheta) = 0,
\end{equation}
where $\oint_{\partial \Theta}$ is a surface integral over $\partial\Theta$, $\vect{n}(\vtheta)$ is the unit vector orthogonal to $\vtheta$ at the boundary $\partial \Theta$ and $S(\mathrm{d}\vtheta)$ is the surface element at $\vtheta \in \partial \Theta$. When $\Theta = \mathbb{R}^d$ is unbounded, condition \eqref{eqn:Boundary} becomes a tail condition which is satisfied if $\int_{\Gamma_r} p(\vtheta) \nabla_{\vtheta} g(\vtheta)\cdot \vect{n}(\vtheta) S(\mathrm{d}\vtheta) \rightarrow 0$ as $r\rightarrow \infty$ where $\Gamma_r \in \mathbb{R}^d$ is a sphere centred at the origin with radius $r$ and $\vect{n}$ is the unit vector orthogonal to $\vtheta$ at $\Gamma_r$. This requirement is given in Equation 9 of \citet{Mira2013} and Assumption 2 of \citet{Oates2017a} and it is a direct result of applying the divergence theorem to $\mathbb{E}_{p}[\mathcal{L}g(\vtheta)]=0$. In the ZV-CV context for $\Theta=\mathbb{R}^d$, a sufficient condition for \eqref{eqn:Boundary} is that the tails of $p$ decay faster than polynomially \citep[Appendix B of][]{Oates2016}.

The unbiased estimator which uses independent samples for estimation of $g(\vtheta)$ and evaluation of \eqref{eqn:MCwithCV} is referred to as the ``split" estimator. In practice, the so-called ``combined'' estimator which uses the full set of $N$ samples for both estimation of $g(\vtheta)$ and evaluation of \eqref{eqn:MCwithCV} can have lower mean square error than the split estimator but is no longer unbiased. If MCMC methods are employed then bias is unavoidable and the combined estimator is likely to be preferred.

\subsection{Parameterization}

An additional consideration when performing either ZV-CV or CF is the adopted parameterization. Any deterministic, invertible transformation of the random variables $\vect{\psi} = f(\vtheta)$ can be used so one can estimate
\begin{align}\label{eqn:MCwithadjCV}
\widehat{\mathbb{E}_{p_{\vtheta}}[\varphi(\vtheta)]} =\frac{1}{N}\sum_{i=1}^N \Big(\varphi(f^{-1}(\vect{\psi}_i)) + \Delta_{\vect{\psi}} g(\vect{\psi}_i)+  \nabla_{\vect{\psi}}g(\vect{\psi}_i) \cdot \nabla_{\vect{\psi}} \log p_{\vect{\psi}}(\vect{\psi}_i)\Big),
\end{align}
instead of \eqref{eqn:MCwithCV}, where $p_{\vtheta}\equiv p$ is the probability density function for $\vtheta$, $p_{\vect{\psi}}$ is the probability density function for $\vect{\psi}$ obtained through a change of measure and $\{\vect{\psi}_i\}_{i=1}^N \sim p_{\vect{\psi}}$. For simplicity, the $\vtheta$ parameterization is used in notation throughout the paper. The best parametrization to adopt for any given application is an open problem. If the original parameterization does not satisfy boundary condition \eqref{eqn:Boundary}, one could consider a reparameterization such that the boundary condition is satisfied.

\section{Regularized Zero-Variance Control Variates}\label{sec:Regularization}

The aim of this section is to develop methods which are computationally less demanding than CF and offer improved statistical efficiency over standard ZV-CV. We describe two types of regularization: regularization through penalized regression and \textit{a priori} regularization. The latter is primarily for cases where not all derivatives of the log target are available or when $N \ll d$.  Combinations of the two regularization ideas are also possible. Methods to choose between control variates are described in Section \ref{ssec:choose}.

\subsection{Regularization Through Penalized Regression}\label{ssec:pen}
As mentioned earlier, the number of regression parameters in ZV-CV grows rapidly with the order $Q$ of the polynomial and with the dimension $d$ of $\vtheta$.  Therefore, the polynomial order that could be considered is limited by the number of samples required to ensure existence of a unique solution to the OLS problem, eliminating the potential reduction that could be achieved using higher order polynomials. In this section, we propose to use penalized regression techniques to help overcome this problem.

In most contexts, using penalized regression reduces variance at the cost of introducing bias. Recall that the conditions for unbiasedness in Section \ref{ssec:Unbiasedness} do not depend on the mechanism for estimating $\vect{\beta}$, as long as the samples used in estimating $\vect{\beta}$ are independent of those used in evaluating \eqref{eqn:MCwithCV}. Thus, the use of penalized regression methods does not introduce bias into ZV-CV.

The regularization methods introduced in Sections \ref{sssec: L2} and \ref{sssec: L1} involve a penalty function on $\vect{\beta}$ so we use standardization for stability and to be able to employ a single $\lambda$. The regression problem becomes:
\begin{align}\label{eqn:standardize}
(\hat{c},\hat{\vect{\beta}_s}) \in \argmin_{\substack{c \in \mathbb{R} \\ \vect{\beta}_s \in \mathbb{R}^{J}}} \frac{1}{N}\sum_{i=1}^N \big[\varphi_s(\vtheta_i) - c + \vect{\beta}_s^\top\vect{x}_s(\vtheta_i)\big]^2 + \lambda \text{PEN}(\vect{\beta}_s),
\end{align}
where the subscript $s$ is in reference to the response and predictors being standardized by their sample mean and standard deviation. Specifically, using the notation $\bar{a} = \frac{1}{N}\sum_{i=1}^N a_i$ and $\sigma_{a} = \sqrt{\sum_{i=1}^N (a_i-\bar{a})^2/(N-1)}$, we have that $\varphi_s(\vtheta_i) = (\varphi(\vtheta_i) - \overline{\varphi})/\sigma_\varphi$, $\vect{x}_s[j](\vtheta_i) = (\vect{x}[j](\vtheta_i) - \overline{\vect{x}[j]})/\sigma_{\vect{x}_j}$ for $j=1,\ldots,J$ and $\vect{\beta}_s$ represents the coefficients on this standardized scale. The estimated coefficients on the original scale are $\hat{\beta}[j] = \hat{\beta}_s[j]\frac{\sigma_\varphi}{\sigma_{x[j]}}$.

The parameter $\lambda$ is chosen to minimize the $k$-fold cross-validation mean square error.

\subsubsection{$\mathcal{L}_2$ Penalization: \emph{PEN}$(g)=\|\mathbf{\beta}_s\|_2^2$}\label{sssec: L2}

The first type of penalization that we consider is Tikhonov regularization \citep{Tikhonov2013}, or ridge regression as it is known when applied in regression \citep{Hoerl1970}. This involves using a squared $\mathcal{L}_2$ penalty, $\text{PEN}(g)=\|\vect{\beta}_s\|_2^2$. Ridge regression mitigates overfitting and allows for estimation when the regression problem is ill-posed due to a small number of observations. Closed form solutions for $\hat{c}$ and $\hat{\vect{\beta}}$ are available, leading to the same computational cost as OLS of $\mathcal{O}(J^3 + NJ^2)$. The use of $\mathcal{L}_2$ penalized ZV-CV can also be motivated using the results of \citet{Belkin2019}, who argued that using $J \geq N$ with an interpolation-based approach (i.e.\ CF or regularized ZV-CV) can lead to better mean square loss compared to restricting to $J \leq N$ (i.e.\ standard ZV-CV) in situations where there is no reason to pre-suppose the first $J$ basis functions are also the most useful. The latter condition may be satisfied when $\varphi$ is too complex to be well-approximated using control variates based on low order polynomials.

To motivate this particular form of penalization, we now consider interpreting this method as a computationally efficient variant of CF. To facilitate a comparison with the approach of  \citet{Barp2018}, we consider a particular instance of CF with a reproducing kernel Hilbert space $\mathcal{H}$ that is carefully selected to lead to an algorithm with lower computational cost.
Namely, we select a polynomial kernel
$$
k(\bm{\theta},\bm{\theta'}) = \sum_{j=1}^{J} P_j(\bm{\theta}) P_j(\bm{\theta}'),
$$
where $P_j(\bm{\theta})$ denotes the $j$th of all $J$ monomial terms in $\bm{\theta}$ up to order $Q$.
For such a kernel, a well-defined Hilbert space $\mathcal{H} = \text{span}\{P_j\}_{j=1,\dots,J}$ is reproduced and we have an explicit expression for the Hilbert norm
$$
\left\| \sum_{j=1}^{J} \beta_j P_j \right\|_{\mathcal{H}} = \left(  \sum_{j=1}^{J} \beta_j^2 \right)^{1/2},
$$
which reveals the method of \citet{Barp2018} as an $\mathcal{L}_2$-penalized regression method.
As such, the optimization problem in ZV-CV with $\text{PEN}(g)=\|\vect{\beta}_s\|_2^2$ is equivalent to the optimization problem in CF (without the standardization of the response and predictors) and it can be solved as a least-squares problem with complexity $\mathcal{O}(J^3 + NJ^2)$.
The first main contribution of our work is to propose a more practical alternative to the method of  \citet{Barp2018}, which we recall has $\mathcal{O}(N^3)$ computational cost, by using such a finite-dimensional polynomial kernel.
Our results in this direction are empirical (only) and we explore the properties of this method for various values of $Q$ in Section \ref{sec:Examples}.

Tikhonov regularization has been applied implicitly in the context of CF but, to the best of our knowledge, this is the first time that general penalized regression methods have been proposed in the context of ZV-CV. Results in Section \ref{sec:Examples} demonstrate that the new estimators can offer substantial variance reduction in practice when the number of samples is small relative to the number of coefficients being estimated.

\subsubsection{$\mathcal{L}_1$-Penalization: \emph{PEN}$(g)=\|\mathbf{\beta}_s\|_1$}\label{sssec: L1} 

The principal aim in the design of a control variate $h$ is to accurately predict the value that the function $\varphi$ takes at an input $\bm{\theta}^*$ not included in the training dataset $\{(\bm{\theta}_i,\varphi(\bm{\theta}_i))\}_{i=1}^N$.
It is well-understood that $\mathcal{L}_1$-regularization can outperform $\mathcal{L}_2$-regula\-ri\-za\-tion in the predictive context when the function $\varphi$ can be well-approximated by a relatively sparse linear combination of predictors. In our case, the unstandardized predictors are the functions in the set $\{1\} \cup \{\mathcal{L}P_j\}_{ j=1,\dots,J}$. Given that low-order polynomial approximation can often work well for integrands $\varphi$ of interest, it seems plausible that $\mathcal{L}_1$-regularization could offer an improvement over the $\mathcal{L}_2$-regularization used in \citet{Oates2017a,Barp2018}.
Investigating this question is the second main contribution of our work.

In the context of ZV-CV, $\mathcal{L}_1$-penalization can be interpreted as using the least absolute shrinkage and selection operator (LASSO, \citet{Tibshirani1996}). LASSO introduces an $\mathcal{L}_1$ penalty $\text{PEN}(g)=\|\vect{\beta}_s\|_1$ where $||\vect{\beta}_s||_1=\sum_j |\beta_s[j]|$. The effect of the penalty is that some coefficients are estimated to be exactly zero.

\subsection{\textit{A priori} Regularization}\label{ssec:apriori}
As an alternative to penalized regression methods, in this section we consider restricting the function $g$ to vary only in a lower-dimensional subspace of the domain $\Theta \subseteq \mathbb{R}^d$. More specifically, a subset of parameters $S \subseteq \{1,\ldots,d\}$ is selected prior to estimation and the function $g$ is defined, in a slight abuse of notation, as $g(\vtheta)= P(\vtheta[S])$. The log target derivatives, $\nabla_{\vtheta} \log{p(\vtheta)}$, only appear in the control variates \eqref{eqn:Stein} through the dot product $\nabla_{\vtheta} g(\vtheta) \cdot \nabla_{\vtheta} \log{p(\vtheta)}$. Therefore if $j \notin S$ then the derivative $\nabla_{\vtheta[j]} \log{p(\vtheta)}$ is not required. We refer to this approach as \textit{a priori} regularization. 

\textit{A priori} regularization makes ZV-CV feasible when some derivatives cannot be used, for example due to intractability, numerical instability, computational expense or storage constraints. An example of where some derivatives may be difficult to obtain is in Bayesian inference for ordinary differential equation (ODE) models. Evaluating $\nabla_{\vtheta} \log{p(\vtheta)}$ requires the sensitivities of the ODE to be computed, which involves augmenting the system of ODEs with additional equations. If some additional equations render the system stiff, then more costly implicit numerical solvers need to be used and in such cases it would be useful to avoid including sensitivites corresponding to the difficult elements of $\vtheta$. It may also be infeasible to use the $\mathcal{O}(d)$ storage required to run standard ZV-CV. Storing a subset of the parameters and derivatives for use in \textit{a priori} regularization may, however, be achievable. Another benefit of \textit{a priori} ZV-CV is that it reduces the number of coefficients to estimate, making estimation feasible when $N \ll d$. \citet{Zhuo2018} consider similar ideas to \textit{a priori} ZV-CV in the context of Stein variational gradient descent, where they use the conditional independence in $p(\vtheta)$ for probabilistic graphical models to separate high dimensional inference problems into a series of lower dimensional problems. 

The downside of using \textit{a priori} ZV-CV is that the potential for variance reduction is reduced, except for under both conditions (a) $\vtheta[S]$ is independent of $\vtheta[\bar{S}]$ according to $p(\vtheta)$, where $\bar{S} = \{1,\ldots,d\} \setminus S$, and (b) $\varphi(\vtheta)=\varphi(\vtheta[S])$. Outside of this situation, restricting the polynomial to $g(\vtheta)=P(\vtheta[S])$ will give varying levels of performance depending on the subset that is selected. Intuitively, one may wish to choose the subset of variables so that $\vtheta[S]$ and/or $\nabla_{\vtheta[S]} \log{p(\vtheta)}$ have high correlations with $\varphi(\vtheta)$. In practice, this is easiest to do when there is \textit{a priori} knowledge and therefore not all derivatives need to be calculated and stored. Given (b), it is suspected that this method will be more useful for individual parameter expectations than for expectations of functions of multiple parameters.

Estimators using this approach are unbiased under the same conditions as ZV-CV and penalized ZV-CV. This method is also applicable to CF, though nonlinear approximation may be more difficult in this non-parametric setting.

\subsection{Automatic Selection of Control Variates}\label{ssec:choose}
The performance of regularized ZV-CV depends upon the polynomial order, the penalization type and on $S$. We demonstrate in Section \ref{sec:Examples} that the common practice of defaulting to $Q=2$ with OLS is often sub-optimal and also that the optimal control variate depends on a variety of factors including $N$ and $p(\vtheta)$. It has previously been proposed to increase the number of control variates as the sample size increases (see e.g.\ \citet{Portier2018} and the appendices of \citet{Oates2017a}). However, in these existing works the mechanism whereby the complexity of the control variate was increased was not data-dependent.

To choose between control variates in this work, we use 2-fold cross-validation so that our selection is data-dependent. For each combination of penalization type and $S$, we start with polynomial order $Q=1$ and we continue to increase the polynomial order until the average cross-validation error is larger for $Q+1$ than for $Q$. The combination of regularization method and polynomial order which gives the minimum cross-validation error is selected and we perform estimation using that method on the full set of samples. The cross-validation error that we use here is the sums of square residuals in the hold-out set, averaged across the two folds.

\section{Empirical Assessment} \label{sec:Examples}
In this section, we perform comparisons of regularized ZV-CV to ZV-CV and CF on Bayesian inference examples. In Bayesian statistics, the posterior distribution of the parameters $\vtheta$ of a statistical model given observed data $\vect{y}$ is
\begin{equation*}
p(\vtheta|\vect{y}) = \frac{\ell(\vect{y}|\vtheta)p_0(\vtheta)}{Z},
\end{equation*}
where the function $\ell(\vect{y}|\vtheta)$ is the likelihood function, $p_0(\vtheta)$ encorporates prior information and $Z$ is a normalising constant. Interest is in estimating posterior expectations $\int_{\Theta} \varphi(\vtheta)p(\vtheta|\vect{y})\mathrm{d} \vtheta$ and the normalising constant or so-called ``evidence" $Z=\int_{\Theta} \ell(\vect{y}|\vtheta)p_0(\vtheta) \mathrm{d} \vtheta$ for Bayesian model choice. Posterior expectations and $Z$ are typically analytically intractable and challenging to estimate due to the potentially high dimensional integration required.

ZV-CV and CF have both been applied in the context of estimating posterior expectations, for example by \citet{Mira2013,Papamarkou2014,Friel2016,Oates2017a,Baker2017}. \citet{Oates2016} and \citet{Oates2017a} have also applied ZV-CV and CF, respectively, to a thermodynamic integration \citep{Gelman1998,Ogata1989,Friel2008} estimator for the evidence, calling the resulting method controlled thermodynamic integration (CTI). The thermodynamic integration estimator gives the log evidence as the sum of multiple expectations with respect to $p_t$ where $p_t=\ell(\vect{y}|\vtheta)^{t} p_0(\vtheta)/Z_t$ and $t$ is referred to as the inverse temperature. \citet{Oates2017a} use population Monte Carlo \citep{Jasra2007} to obtain the samples from $p_t$ for $t=0,\ldots,T$ and they consider specifically $t_j=(j/T)^5$. A total of $2(T+1)$ expectations are involved, with ZV-CV applied to the estimator for each expectation.

We propose to use sequential Monte Carlo (SMC, \citet{DelMoral2006}) with the tuning method of \citet{Salomone2018} for sampling, rather than the standard choices of MCMC or population MCMC. The benefit of this approach is that the samples are roughly independent which can be preferable over the high autocorrelation that can be seen in MCMC samples. The standard SMC evidence estimator is the product of $T$ expectations, so we consider improving this estimator using ZV-CV and CF. Further details about implementation in SMC and the advantages of this approach are given in Appendix B of the Online Resources. From the perspective of comparing variance reduction methods, the application of ZV-CV to posterior expectations and to multiple evidence estimators means that ZV-CV can be compared on a variety of functions $\varphi(\vtheta)$ and distributions $p(\vtheta)$.

We perform an empirical comparison of the following methods using examples of varying complexity:
\begin{itemize}
\item \textbf{vanilla}: Monte Carlo integration without control variates.
\item \textbf{\zv{Q}}: ZV-CV with OLS and order $Q$ polynomial.
\item \textbf{\zvl{Q}}: ZV-CV with LASSO and order $Q$ polynomial.
\item \textbf{\zvr{Q}}: ZV-CV with ridge regression and order $Q$ polynomial.
\item \textbf{sub\textsubscript{$k$}-}: This prefix indicates \textit{a priori} ZV-CV with a subset of size $k$. Applications are limited to $d>1$ dimensions. We only apply \textbf{sub\textsubscript{$k$}-} ideas to posterior expectations since $\varphi(\vtheta)$ is a function of a single parameter and a potentially reasonable subset may be known \textit{a priori}.
\item \textbf{crossval}: Control variate selection using 2-fold cross-validation. This method chooses between \zv{\text{Q}}, \zvl{\text{Q}}, \zvr{\text{Q}} and \textbf{sub\textsubscript{$k$}-} where applicable.
\item \textbf{CF}: Control functionals with a second-order Stein operator, a Gaussian kernel $k(\vtheta,\vtheta') = \exp(-\|\vtheta - \vtheta'\|_2^2 /\sigma^2)$ and selection of $\sigma^2$ using $5$-fold cross-validation with the generous 15-value grid $10^{\vect{\kappa}}$ where $\kappa_i = -3 + 0.5i$ for $i=0,\ldots,14$.
\end{itemize}
Methods written with the prefix sub\textsubscript{$k$}-  or the name \zvl{Q}, \zvr{Q} or crossval are novel for all $Q$ and $k$. The main purposes of these comparisons are to investigate the performance of higher order polynomials, the utility of penalized regression and the ability to achieve variance reduction using a subset of derivatives. The purpose of the comparisons to CF is not necessarily to outperform CF, as CF can be infeasible to apply in its basic form for large $N$, but to benchmark the performance of these novel methods against CF. A variety of sample sizes, integrands $\varphi(\vtheta)$ and target distributions $p$ are used for fair comparisons. We focus on sample sizes that are typical of SMC, ranging from $N=10$ to $N=10,000$ but we note that larger sample sizes can be accommodated by the regularized ZV-CV methods which have a computational complexity of $\mathcal{O}(N)$. 

Estimators are compared on the basis of mean square error (MSE), where the gold standard of estimation is carefully chosen for each example. The main quantity of interest reported in this section is $\widehat{\mathsf{MSE}_p[\text{vanilla}]}/\widehat{\mathsf{MSE}_p[\cdot]}$, the MSE of the vanilla Monte Carlo estimator estimated from 100 independent SMC runs divided by the estimated MSE for the method in question. This quantity is referred to as statistical efficiency and it is reported for each fixed $N$. Values above one are preferred.

Control variate methods are most valuable when the sampling algorithm is expensive, for example due to the cost of evaluating the likelihood, or when evaluation of the function $\varphi(\vtheta)$ is costly. The overall efficiency, as measured by
\begin{equation*}
\frac{\widehat{\mathsf{MSE}_p[\text{vanilla}]} \times \widehat{\mathsf{time}[\text{vanilla}]}} {\widehat{\mathsf{MSE}_p[\cdot]}  \times  \widehat{\mathsf{time}[\cdot]}},
\end{equation*}
is also considered for these examples. Here $\widehat{\mathsf{time}[\cdot]}$ is the average time across the 100 runs to compute the estimator in question, including the time spent running the SMC sampler. We note that the run time is subject to the efficiency of the code and here (penalized) ZV-CV is based on the R package glmnet, cross-validated ZV-CV is written as a loop in R and CF is implemented in $\text{C}$\texttt{++}. Nevertheless, our proposed methods offer improved overall efficiency in several of the applications considered. The computational benefits of our approach will improve with increasing model complexity in terms of likelihood calculations, since the overhead associated with penalized regression will become relatively negligible.

Two examples are described in detail in this section. Appendices E, F and G also include results for a 61-dimensional logistic regression example, a one-dimensional ODE example which motivates higher order polynomials and a challenging nine-dimensional ODE model, respectively.

In terms of bias, boundary condition \eqref{eqn:Boundary} is satisfied using the specified parameterizations for all examples considered in this paper. This can be verified through the sufficient condition that the tails of $p$ decay faster than polynomially and $\Theta = \mathbb{R}^d$ (Appendix B of \citet{Oates2016}). However, the estimators are generically biased due to the use of SMC, as they would be with MCMC. All results are based on combined estimators as opposed to split estimators, so all pairs $\{\vtheta_i,\varphi(\vtheta_i)\}_{i=1}^N$ are used to build $\tilde{\varphi}$ and also to estimate $\mathbb{E}_{p}[\tilde{\varphi}(\vtheta)]$.

\subsection{Recapture Example}\label{ssec:recapture}
This 11-dimensional example demonstrates that reduced variance estimators can be obtained with the use of higher order polynomials and regularization.

\citet{Marzolin1988} collected data on the capture and recapture of the bird species \textit{Cinclus cinclus}
over six years. Like \citet{Brooks2000}, \citet{Nott2018} and \citet{South2018}, we use a Bayesian approach to estimate the parameters of a Cormack-Jolly-Seber model \citep{Lebreton1992} for the capture and recapture of this species. The parameters of the Cormack-Jolly-Seber model used here are the probability of survival from year $i$ to $i+1$, $\phi_i$, and the probability of being captured in year $k$, $p_k$, where $i=1,\ldots,6$ and $k=2,\ldots,7$. Denote the number of birds released in year $i$ as $D_i$ and the number of animals caught in year $k$ out of the number released in year $i$ as $y_{ik}$. It is simple to show that the number released in year $i$ that are never caught is $d_i=D_i-\sum_{k=i+1}^{7}y_{ik}$ and the probability of a bird being released in year $i$ and never being caught is $\chi_i=1-\sum_{k=i+1}^{7} \phi_i p_k \prod_{m=i+1}^{k-1} \phi_m (1-p_m)$. The likelihood is given by
\begin{align*}
\ell(\vect{y}|\mathbf{\vtheta}) \propto \prod_{i=1}^{6}\chi_i^{d_i} \prod_{k=i+1}^{7} \left[ \phi_i p_k \prod_{m=i+1}^{k-1} \phi_m (1-p_m) \right]^{y_{ik}},
\end{align*}
where $\vtheta = (\phi_1,\ldots,\phi_5,p_2,\ldots,p_6,\phi_6p_7)$. Following \citet{South2018}, the parameters $\phi_6$ and $p_7$ are multiplied together due to a parameter identifiability issue.

The prior is $\vtheta[j] \sim \mathcal{U}(0,1)$ for $j=1,\ldots,11$. To satisfy the boundary conditon \eqref{eqn:Boundary} and to improve the efficiency of MCMC proposals, all parameters are transformed to the real line using $\vect{\psi}[j]=\log(\vtheta[j]/(1-\vtheta[j]))$ so the prior density for $\vect{\psi}[j]$ is $\exp(\vect{\psi}[j])/(1+\exp(\vect{\psi}[j]))^2$, for $j=1,\ldots,11$.

The gold standard of evidence estimation for this example is the mean evidence estimate for \zvl{1} at $N=5000$. The posterior expectation gold standard is the average posterior mean for \zv{4} at $N=5000$.

\subsubsection{Posterior Expectations}
\begin{figure*}
\centering
\subcaptionbox{Statistical efficiency}{\includegraphics[width=0.48\textwidth]{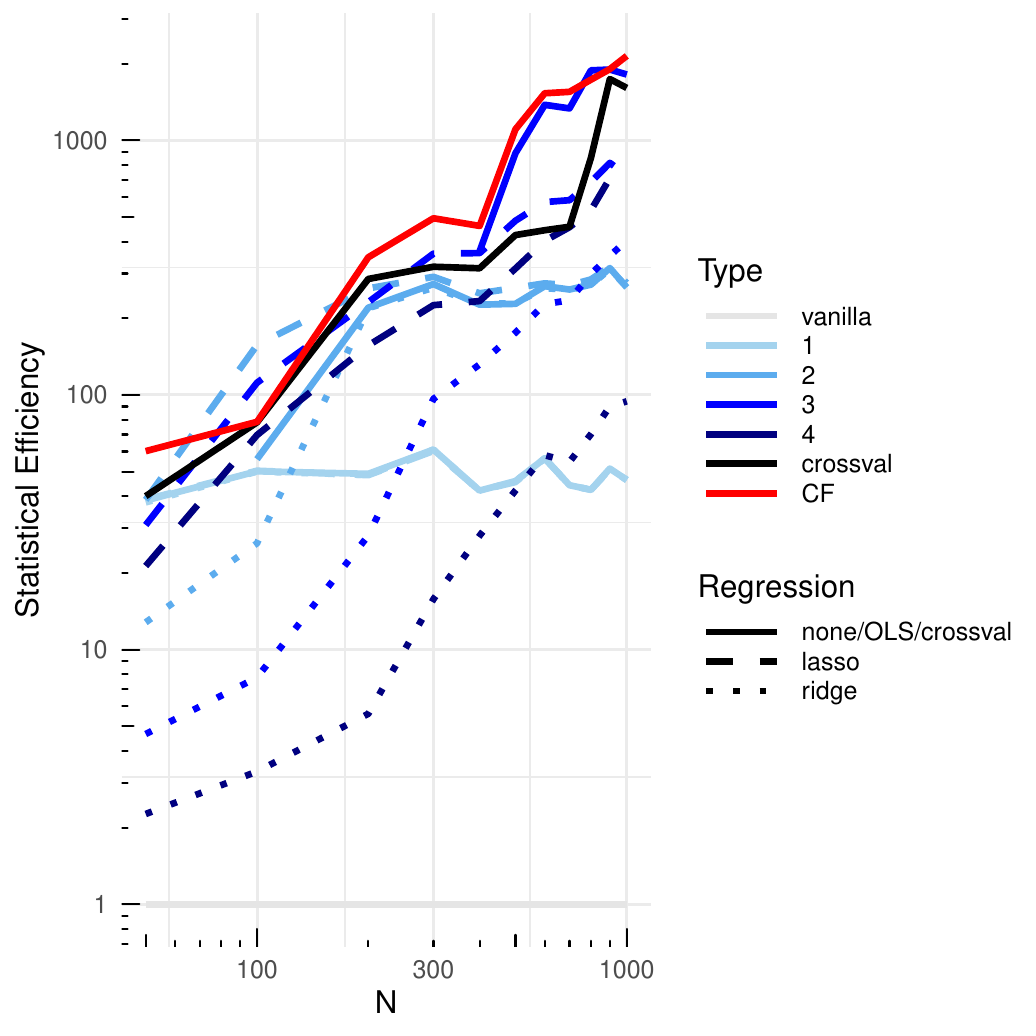}}
\subcaptionbox{Overall efficiency}{\includegraphics[width=0.48\textwidth]{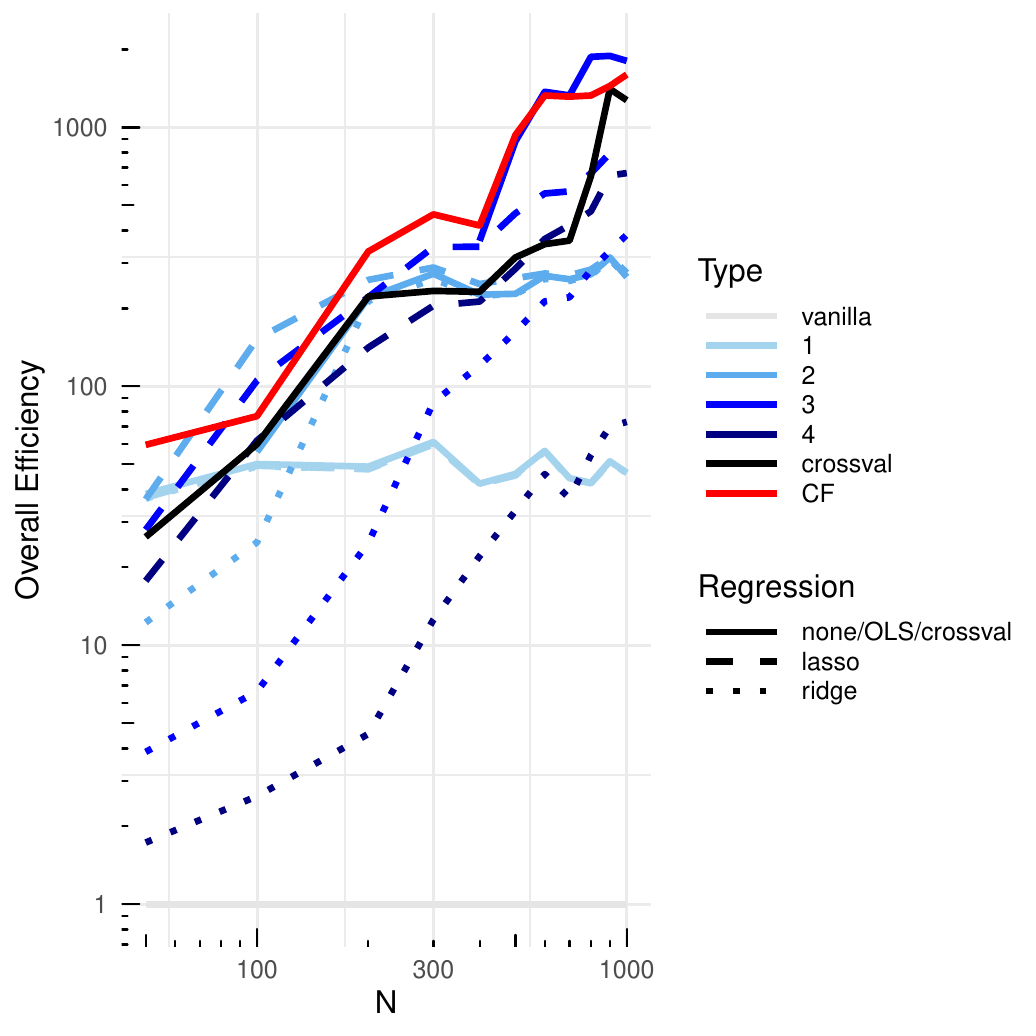}}
\caption{Recapture example: (a) statistical efficiency averaged over 11 parameters and (b) overall efficiency averaged over 11 parameters.}
\label{fig:recapture_posterior_both}
\end{figure*}

The average statistical efficiency and overall efficiency across parameters is shown in Figure \ref{fig:recapture_posterior_both}, excluding \textit{a priori} regularization results for simplicity. Higher order polynomials become more efficient as $N$ increases and the use of penalized regression means that higher order polynomials can be considered for smaller $N$. LASSO regression is preferable over ridge regression for this example. 

Using \textit{a priori} ZV-CV with $S=j$, where $j$ is the index of the current parameter of interest, \zvs{1}{1} is on average roughly 10 times more efficient than vanilla Monte Carlo integration. 

Cross-validation generally gives similar results to CF and to the best performing fixed method. More details about the selected control variates can be found in Appendix C of the Online Resources.

\subsubsection{Evidence Estimation}

\begin{figure}[!h]
\centering
\subcaptionbox{Statistical Efficiency CTI}{\includegraphics[width=0.4\textwidth]{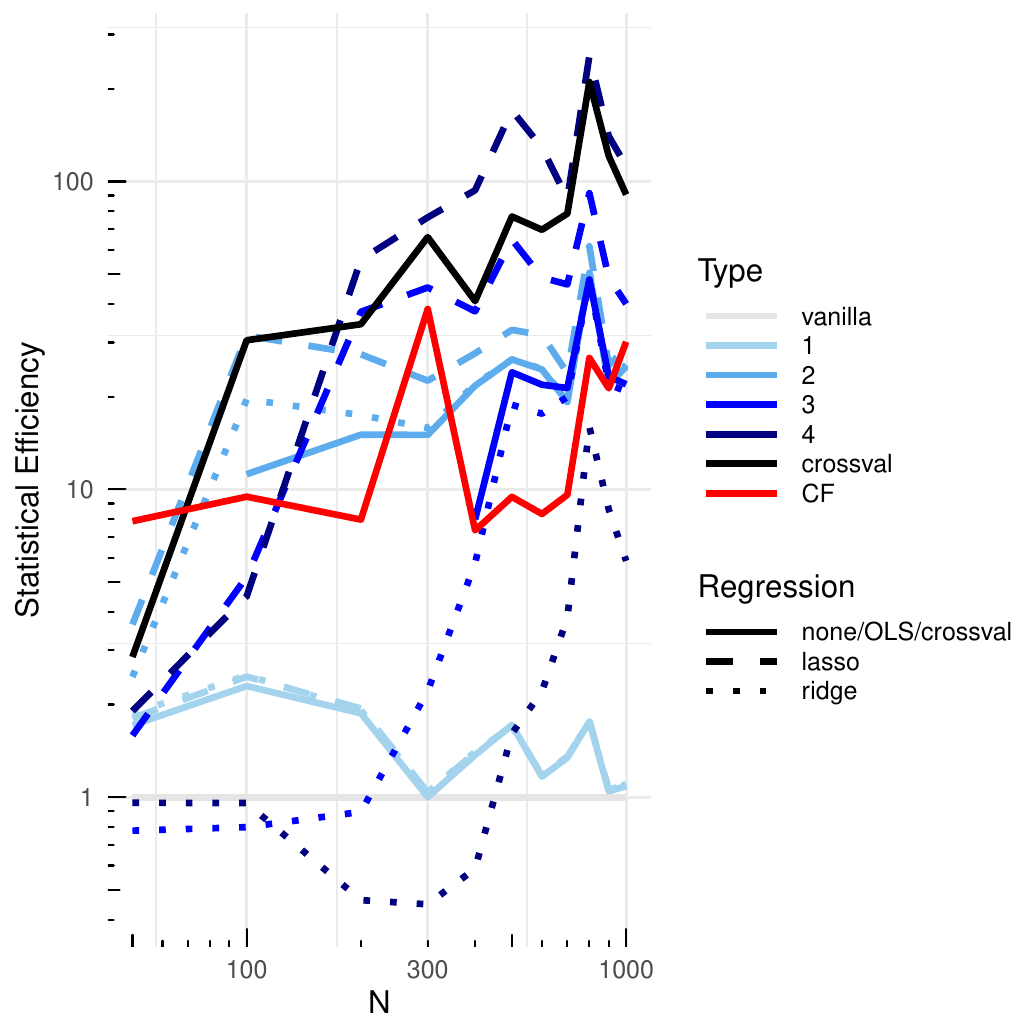}}
\subcaptionbox{Overall Efficiency CTI}{\includegraphics[width=0.4\textwidth]{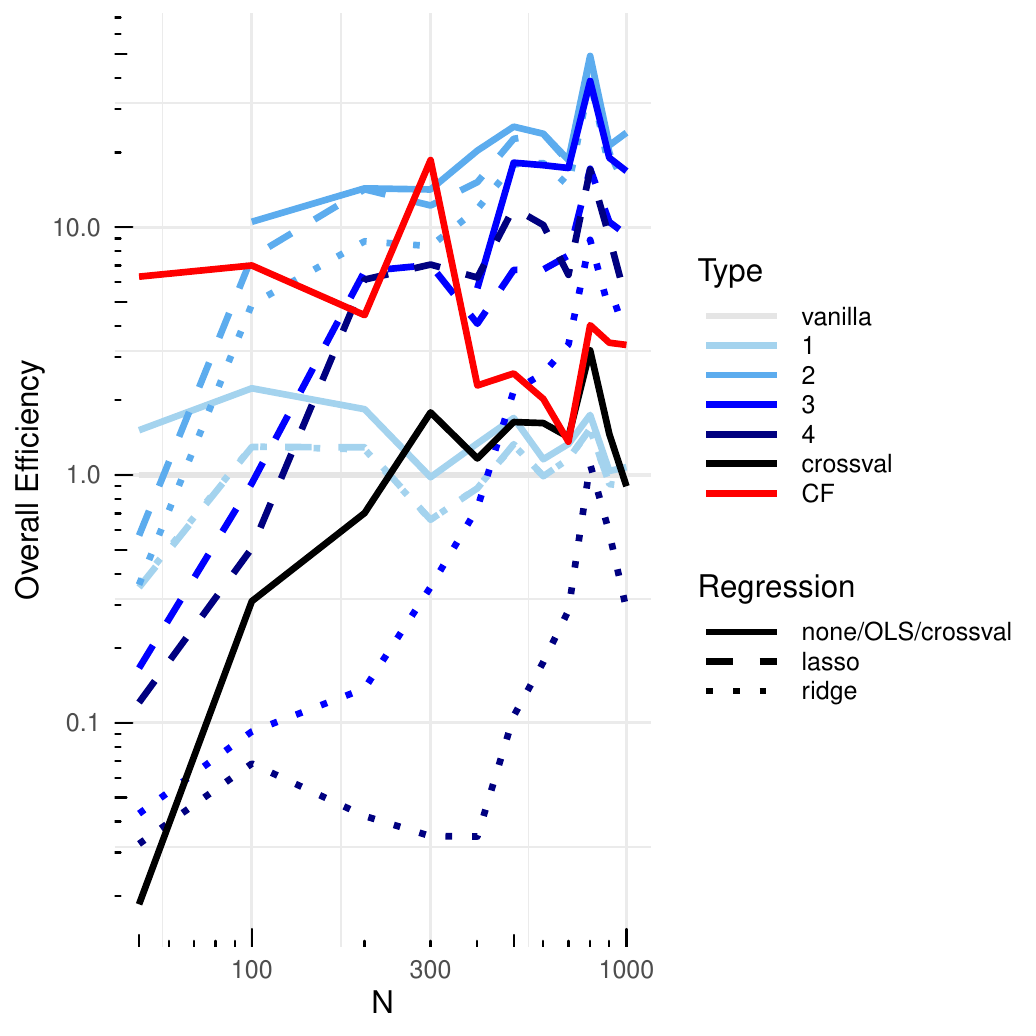}}
\subcaptionbox{Statistical Efficiency SMC}{\includegraphics[width=0.4\textwidth]{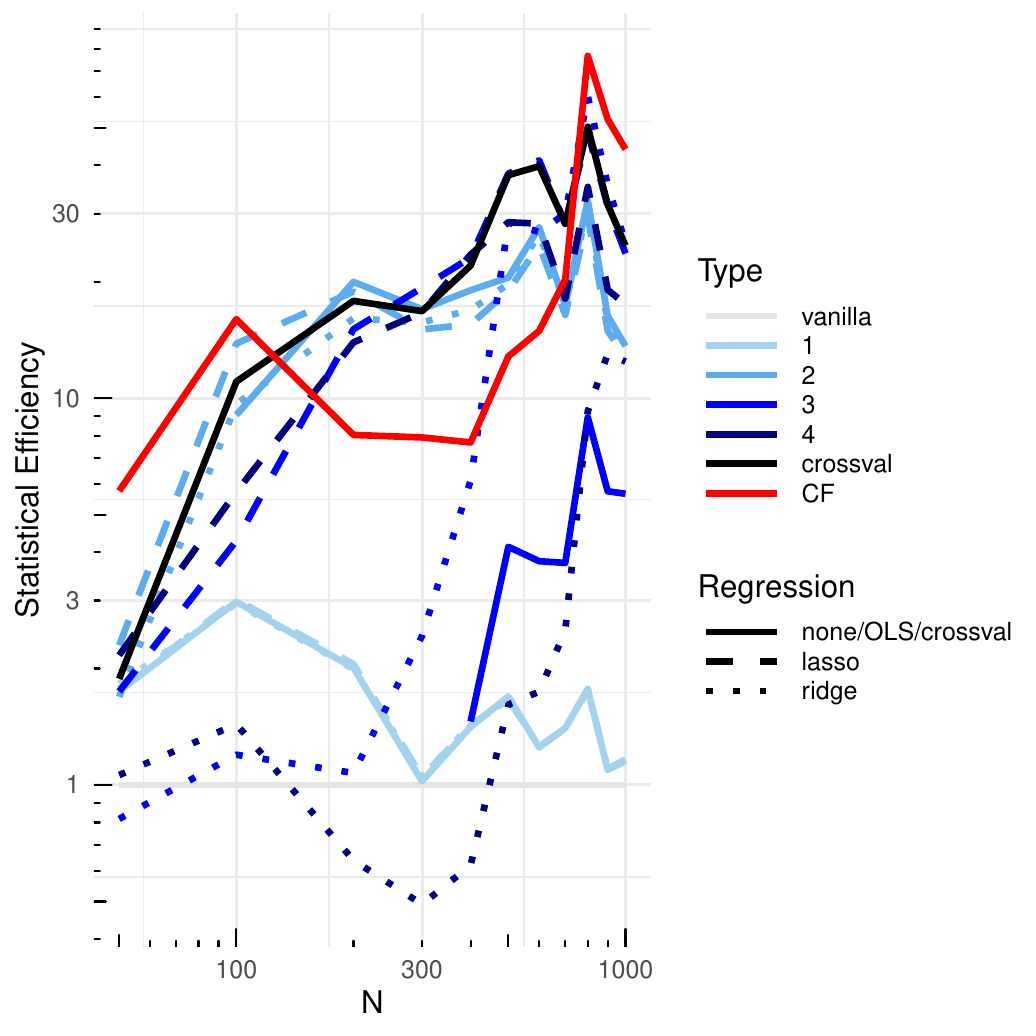}}
\subcaptionbox{Overall Efficiency SMC}{\includegraphics[width=0.4\textwidth]{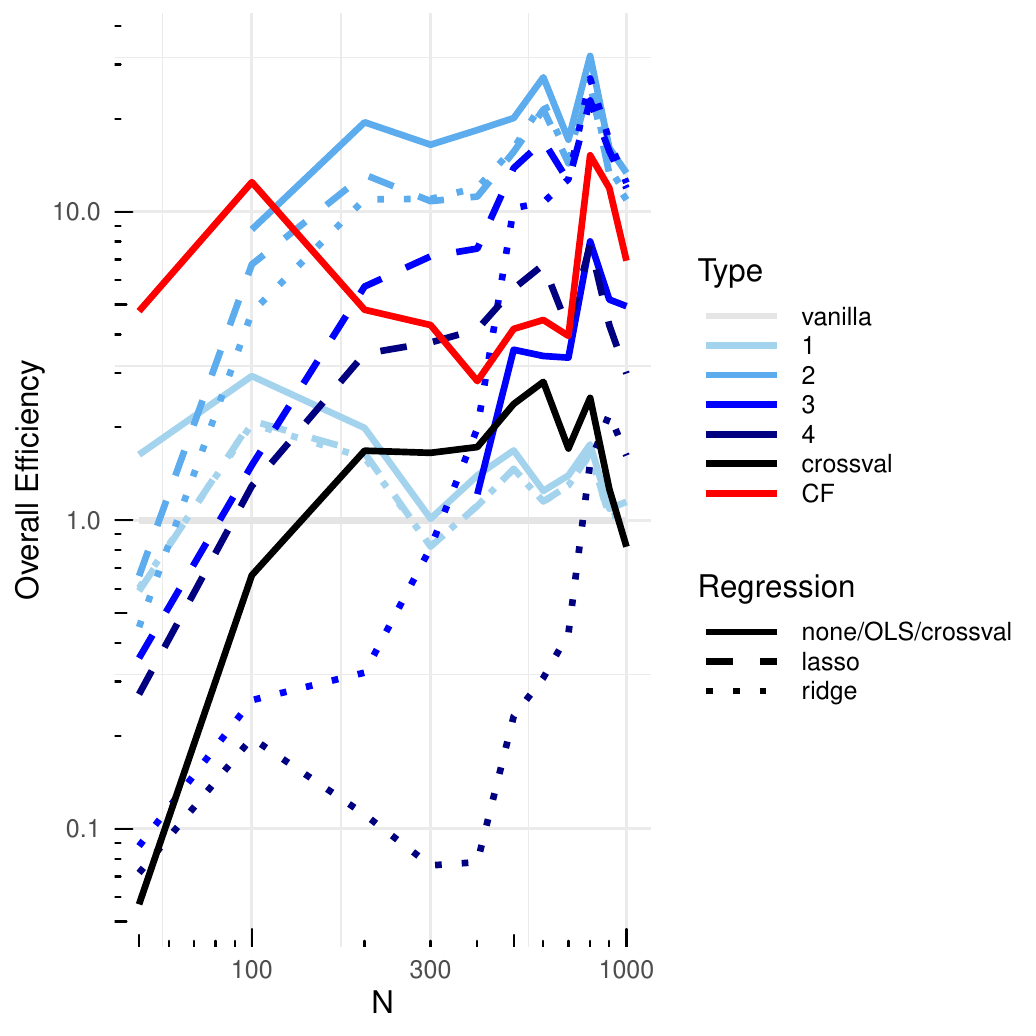}}
\caption{Recapture example: (a) statistical efficiency for the CTI estimator, (b) overall efficiency for the CTI estimator, (c) statistical efficiency for the SMC estimator and (d) overall efficiency for the SMC estimator.}
\label{fig:recapture_evidence}
\end{figure}

Regularized ZV-CV and automatic control variates give improved statistical efficiency over ZV-CV and CF for the range of $N$ that are considered here, as seen in Figure \ref{fig:recapture_evidence}. However, there is less improvement in terms of overall efficiency due to the fact that multiple expectations are required for evidence estimation. This puts the more computationally intensive methods including higher order polynomials, cross-validation and CF at a significant disadvantage. We note that this example was selected to allow for extensive comparisons and the cost of post-processing would have less impact under more expensive likelihood functions.

The selected control variates for $N=50$ and $N=1000$ can be found in Appendix C of the Online Resources. 

\subsection{Log-Gaussian Cox Point Process Example}\label{ssec:cox}
We now consider an example where the dimension can be adjusted. The log-Gaussian Cox point process example of \citet{Moller1998} consists of locations of 126 Scots pine saplings in a $10 \times 10$ m$^2$ plot. The plot can be discretised into $n \times n$ grid cells, so that the dimension $d=n^2$ of the problem can be varied. Here we consider $n=4$, $n=8$ and $n=16$ so that we have Bayesian inference problems of size $d=16$, $d=64$ and $d=256$. 

The model specifications, including code for the log likelihood, log prior and their gradients, match that of \citet{Heng2019}. After normalising the plot to fit onto a unit square, the number of points at grid cell $(i,j)$ for $i,j = 1,\ldots,n$ is denoted $y_{i,j}$. It is assumed that the $y_{i,j}$ are conditionally independent and Poisson distributed with mean $\frac{\Lambda_{i,j}}{n^2}$. The prior is $\Lambda_{i,j} = \exp(\theta_{i,j})$ where $\theta_{i,j}$ has a Gaussian process prior with mean $\mu$ and covariance function $\Sigma_{(i,j),(i',j')}=s^2 \exp \left[ - \sqrt{(i-i')^2+(j-j')^2}/(n\beta)\right]$, where $\beta = 1/33$, $s^2 = 1.91$ and $\mu = \log(126) - s^2/2$.

Our goal is to estimate the posterior means for the parameters $\theta_{i,j}$ for $i,j=1,\ldots,n$, and we do so using SMC runs with $N=100$ particles. We do not consider evidence estimation due to lack of a reliable gold standard. Due to memory and time constraints, the maximum polynomial order is constrained so that the maximum number of covariates in ZV-CV is 5000. The gold standards in this example are the average posterior expectations across many independent unbiased Riemann-manifold HMC runs \citep{Heng2019} with unbiased control variates as described in \citet{South2019discussion}. Details of the gold standard are available in Appendix D.

Tables \ref{tab:cox16_100}, \ref{tab:cox64_100} and \ref{tab:cox256_100} show the mean relative statistical, computational and overall efficiency for posterior expectations in dimensions $d=16$, $d=64$ and $d=256$, respectively, when $N=100$. In all three settings, the existing methods (vanilla MC, ZV-CV with OLS and CF) are outperformed by the novel approaches of \textit{a priori} ZV-CV, cross-validation and LASSO with a higher order polynomial than OLS could handle. The best performing novel method has an overall efficiency which is better than the best performing existing method by a factor of over 1,600,000 for $d=16$, over 1,800 for $d=64$ and over 25 for $d=256$. Results showing the competitive performance of the novel methods for $N=1,000$ and $N=10,000$ are given in Appendix D. Like the results for $N=100$, \textit{a priori} ZV-CV, cross-validation and LASSO outperform existing alternatives in the majority of settings.


\begin{table*}
\centering
\caption{16-dimensional Cox example: efficiency for marginal posterior expectations when $N=100$, averaged over results for all 16 parameters. A ``$-$''' indicates that the population size $N=100$ is insufficient for standard ZV-CV. We refer the reader to the beginning of Section 4 for acronym definitions.} 
\label{tab:cox16_100}
\begin{tabular}{llrrr}
Approach & $Q$ & Stat. Efficiency & Comp. Efficiency & Overall Efficiency\\ \hline
Vanilla & NA & $1.0 \times 10^0$ & $1.0000$ & $1.0 \times 10^0$\\
CF & NA & $4.4 \times 10^1$ & $0.9787$ & $4.3 \times 10^1$\\ \hline
\zv{1} & 1 & $2.0 \times 10^1$ & $0.9994$ & $2.0 \times 10^1$ \\
\zvl{1} & 1 & $2.2 \times 10^1$ & $0.9722$ & $2.2 \times 10^1$ \\
\zvr{1} & 1 & $2.0 \times 10^1$ & $0.9719$ & $1.9 \times 10^1$ \\ \hline
\zv{2} & 2 & $-$ & $-$ & $-$ \\
\zvl{2} & 2 & $1.2\times10^3$ & $0.9315$ & $1.1 \times 10^3$ \\
\zvr{2} & 2 & $2.1\times10^1$ & $0.8435$ & $1.8 \times 10^1$ \\ \hline
\zv{3} & 3 & $-$ & $-$ & $-$ \\
\zvl{3} & 3 & $5.0\times 10^2$ & $0.8330$ & $4.2 \times 10^2$\\
\zvr{3} & 3 & $1.7\times 10^1$ & $0.6675$ & $1.1 \times 10^1$ \\ \hline
\zv{4} & 4 & $-$ & $-$ & $-$ \\
\zvl{4} & 4 & $3.2\times 10^2$ & $0.6151$ & $2.0 \times 10^2$\\
\zvr{4} & 4 & $1.4 \times 10^1$ & $0.3857$ & $5.3 \times 10^0$ \\ \hline
\zvs{1}{1} & 1 & $2.4 \times 10^1$ & $0.9996$ & $2.4 \times 10^1$ \\
\zvs{2}{1} & 2 & $5.0 \times 10^3$ & $0.9996$ & $5.0 \times 10^3$ \\
\zvs{3}{1} & 3 & $3.5 \times 10^6$ & $0.9996$ & $3.5 \times 10^6$ \\
\zvs{4}{1} & 4 & $6.9 \times 10^7$ & $0.9995$ & $6.9 \times 10^7$ \\ \hline
crossval & NA & $2.8 \times 10^8$ & $0.2125$ & $5.5 \times 10^7$\\
\end{tabular}
\normalsize
\end{table*}

\begin{table*}
\centering
\caption{64-dimensional Cox example: efficiency for marginal posterior expectations when $N=100$, averaged over results for all 64 parameters. A ``$-$''' indicates that the population size $N=100$ is insufficient for standard ZV-CV. We refer the reader to the beginning of Section 4 for acronym definitions.} 
\label{tab:cox64_100}
\begin{tabular}{llrrr}
Approach & $Q$ & Stat. Efficiency & Comp. Efficiency & Overall Efficiency\\ \hline
Vanilla & NA & $1.0$ & $1.0000$ & $1.0$\\
CF & NA & $5.9$ & $0.9953$ & $5.9$\\ \hline
\zv{1} & 1 & $4.5$ & $0.9999$ & $4.5$\\
\zvl{1} & 1 & $12.0$ & $0.9931$ & $11.9$\\
\zvr{1} & 1 & $5.2$ & $0.9924$ & $5.2$\\ \hline
\zv{2} & 2 & $-$ & $-$ & $-$\\
\zvl{2} & 2 & $17.2$ & $0.9574$ & $16.5$\\
\zvr{2} & 2 & $5.2$ & $0.9071$ & $4.7$\\ \hline
\zvs{1}{1} & 1 & $13.8$ & $0.9999$ & $13.8$\\
\zvs{2}{1} & 2 & $2505.3$ & $0.9999$ & $2505.1$\\
\zvs{3}{1} & 3 & $7181.6$ & $0.9999$ & $7181.0$\\
\zvs{4}{1} & 4 & $11081.3$ & $0.9999$ & $11080.4$\\ \hline
crossval & NA & $7271.9$ & $0.4631$ & $3369.3$\\
\end{tabular}
\normalsize
\end{table*}

\begin{table*}
\centering
\caption{256-dimensional Cox example: efficiency for marginal posterior expectations when $N=100$, averaged over results for all 256 parameters. A ``$-$''' indicates that the population size $N=100$ is insufficient for standard ZV-CV. We refer the reader to the beginning of Section 4 for acronym definitions.} 
\label{tab:cox256_100}
\begin{tabular}{llrrr}
Approach & $Q$ & Stat. Efficiency & Comp. Efficiency & Overall Efficiency\\ \hline
Vanilla & NA & $1.0$ & $1.0000$ & $1.0$\\
CF & NA & $2.0$ & $0.9990$ & $2.0$\\ \hline
\zv{1} & 1 & $-$ & $-$ & $-$\\
\zvl{1} & 1 & $21.2$ & $0.9989$ & $21.2$\\
\zvr{1} & 1 & $2.1$ & $0.9980$ & $2.1$\\ \hline
\zvs{1}{1} & 1 & $21.1$ & $1.0000$ & $21.1$\\
\zvs{2}{1} & 2 & $52.3$ & $1.0000$ & $52.3$\\
\zvs{3}{1} & 3 & $53.4$ & $1.0000$ & $53.4$\\
\zvs{4}{1} & 4 & $42.3$ & $1.0000$ & $42.3$\\ \hline
crossval & NA & $32.3$ & $0.9855$ & $31.8$\\
\end{tabular}
\normalsize
\end{table*}

\section{Discussion} \label{sec:Discussion}

In this paper, we introduced two types of regularized ZV-CV: regularization through penalized regression and regularization by selecting a subset of parameters to include in the regression model. Higher order polynomial basis functions have the potential to outperform the commonly used polynomial with $Q=2$ as $N$ - the number of Monte Carlo, MCMC or SMC simulations - increases. Our penalized ZV-CV ensures that the resulting functional approximation problem remains well-defined when $N$ is less than the number of control variate coefficients ($J$) while performing similarly to standard ZV-CV when $N > J$. For the examples considered here, we found that LASSO generally resulted in better performance than ridge regression. \textit{A priori} ZV-CV led to significant improvements over vanilla Monte Carlo for posterior expectations, with little computational overhead. 

One of the main applications of the proposed methods is in models where the dimension, $d$, is too high for standard variance reduction techniques to be efficient. Empirical evidence suggests that using ZV-CV and penalized ZV-CV, where $Q$ is increased with $N$, offers better statistical performance than CF in high dimensions. However, the computational cost of (penalized) ZV-CV is $\mathcal{O}(N{{d + Q}\choose{d}}^2 + {{d + Q}\choose{d}}^3)$, which may prohibit the application of these methods with large $Q$ in high dimensions. This explosion in complexity for large $Q$ and $d$ is a disadvantage relative to CF when the sample size is comparable or less than the dimension, though the complexity is similar when $N \approx d$ and $Q=1$ in (penalized) ZV-CV. One could consider speeding up these algorithms by using partial LASSO searches \citep[e.g.][]{Efron2004,Fan2008} or by using approximate solvers as proposed in \citet{Si2020}. Alternatively, in very large dimensions, the \textit{a priori} ZV-CV approach can be used to obtain variance reductions with a complexity that is $\mathcal{O}(N |S|^2 + |S|^3)$ where $1 \leq |S| \leq d$. This \textit{a priori} approach also offers benefits when not all derivatives are available, when $N\ll d$, or when information about the relationships between the integrand and parameters is known (for example when $p(\vtheta)$ has a directed acyclic graph factorization). 

\citet{Leluc2019} provide additional theoretical support for LASSO-based control variate selection. The work of \citet{Leluc2019}, which was publicly available after the pre-print of our paper \citep{South2018Unpublished}, gives concentration inequalities for the integration error with LASSO-based control variates and also shows that the correct control variates are selected with high probability. The theoretical results are based on bounded control variates, which do not apply in ZV-CV and CF when $p$ has unbounded support. \citet{Leluc2019} find empirically that a methodological adjustment of performing OLS for estimation once the control variates have been selected via LASSO is helpful in reducing the variance of the estimator. We point out that this modification is necessary to obtain the zero-variance property of ZV-CV. The optimal coefficients required to obtain zero-variance estimators cannot be obtained directly from penalized regression methods like LASSO and ridge regression with non-zero $\lambda$.

We have proposed the consideration of different penalty functions in the optimization problem for control variates, but we focus specifically on LASSO and ridge regression. Some other potentially useful regularization methods for the situation where $N < {{d + Q}\choose{d}}$ are elastic net \citep{Zou2005} and partial least squares (PLS, \citet{Wold1975}). Elastic net is a compromise between LASSO and ridge regression which uses two tuning parameters. PLS is based on choosing the $k<{{d + Q}\choose{d}}-1$ independent linear combinations of covariates that explain the maximum variance in the response, where $k$ is chosen through cross-validation. Active subspaces \citep{Constantine2015} are a more recent dimension-reduction technique which use the derivatives of the function of interest to find the linear combinations of covariates that are best at predicting the function. It would be of interest in future research to compare our LASSO and ridge regression ZV-CV methods with these alternatives.

The concept of regularization by selecting a subset of parameters is referred to as nonlinear approximation in approximation theory and applied mathematics \citep{Devore1998}, and there is some theoretical evidence to suggest that this can outperform linear approximation (e.g.\ penalized regression which is described in Section \ref{ssec:pen}). Selecting a particular subset of monomials which are used in a polynomial interpolant is also the same idea as in sparse grid algorithms for numerical integration \citep{Smolyak1963}. These methods are known to work well in high dimensions and could be useful alternatives for selecting the subset of monomials in ZV-CV.

Stein-based control variates using neural networks have recently appeared in the literature \citep{Zhu2018}. \citet{Zhu2018} added details of penalization methods to their approach, where the control variates cannot be fitted exactly and stochastic optimization is required. Penalization methods are simpler and more stable in the linear regression context but in future research it would be of interest to compare to neural control variates with regularization. This alternative approach is likely to outperform ZV-CV in some applications, such as when $\varphi(\vtheta)$ is multi-modal.  

Finding the optimal parameterization for a given application is a challenging open problem. Choosing the parameterization is a trade-off between making $p_\phi$ simpler and making $\varphi(f^{-1}(\phi))$ simpler. Another potential benefit of reparameterising for ZV-CV is that there is the potential to enforce more sparsity in the predictors  for improved performance in $\mathcal{L}_1$ penalization.

Derivatives are available in closed form or can be unbiasedly estimated for a large class of problems. ZV-CV has been applied in big data settings in the context of post-processing after stochastic gradient MCMC \citep{Baker2017} and for models with intractable likelihoods \citep{Friel2016}. Regularized ZV-CV also applies in these settings. Regularized ZV-CV could also be used in exact approximate settings where a particle filtering estimate of the likelihood is used (see for example \citet{Dahlin2015} and \citet{Nemeth2016}). However, derivative-based methods are most appealing when the derivative of the log target can be obtained with little additional cost relative to the likelihood itself. An interesting avenue for future research may be to consider automatic differentiation. 

\section*{Acknowledgement}
The authors thank anonymous referees and the associate editor for helpful comments. The authors also wish to thank Nial Friel for the suggestion to reduce the variance of the SMC evidence estimator using ZV-CV and for comments on an earlier draft. LFS and CD are associated with the ARC Centre of Excellence for Mathematical \& Statistical Frontiers (ACEMS). LFS would like to thank Matthew Sutton for useful discussions about penalized regression methods. LFS was supported by an Australian Research Training Program Stipend, by ACEMS and by the Engineering and Physical Sciences Research Council grant EP/S00159X/1. CJO was supported by the Lloyd's Register Foundation programme on data centric engineering at the Alan Turing Institute, UK.  CD and CJO were supported by an Australian Research Council Discovery Project (DP200102101). AM was partially supported by the Swiss National Science Foundation grant 100018\_200557. Computational resources used in this work were provided by the HPC and Research Support Group, Queensland University of Technology, Brisbane, Australia and by the High End Computing facility at Lancaster University.

\bibliographystyle{apalike}
\bibliography{Refs}

\clearpage

\appendix

\section{Covariates in the ZV-CV Regression}\label{app:CV_PolyEstimation}

As described in Section 2.1 of the main paper, ZV-CV uses control variates of the form $\vect{\beta}^{\top}\vect{x}(\vtheta)$ where $\vect{\beta} \in \mathbb{R}^{J}$, $J = {{d + Q}\choose{d}}-1$, $\vtheta \subseteq \mathbb{R}^d$ and $Q$ is the polynomial order. This appendix gives the general form of $\vect{x}(\vtheta)$ for a fixed $Q$.

The $j$th element of $\vect{x}(\vtheta)$ is:
\begin{align*}
\vect{x}[j] &= \sum_{k=1}^d \vect{A}_{j,k} \left[\vtheta[k]^{\vect{A}_{j,k}-1}\nabla_{\vtheta[k]} \log{p(\vtheta)}  + (\vect{A}_{j,k}-1) \vtheta[k]^{\vect{A}_{j,k}-2}\right]\prod_{z=1, z\neq k}^{d} \vtheta[z]^{\vect{A}_{j,z}},
\end{align*}
where $j=1,\ldots,J$. The matrix $\vect{A} \in \mathbb Z_{\geq 0}^{J\times d}$ has $J$ rows where each row corresponds to a unique vector $\vect{A}_{j,\cdot}$ such that $1 \leq \sum_{k=1}^d \vect{A}_{j,k} \leq Q$. In other words, $\vect{A}$ contains all permutations of powers to $\vtheta[1],\ldots,\vtheta[d]$ that lead to a sum of exponents between $1$ and $Q$.

It is straightforward to verify that $\vect{x} = \nabla_{\vtheta} \log{p(\vtheta)} \in  \mathbb{R}^{d} $ for the first order polynomial $P_1(\vtheta) = c + \sum_{k=1}^d \beta[k] \vtheta[k]$.

\section{Variance Reduction in Sequential Monte Carlo}\label{app:SMC}
In this appendix, the novel applications of Stein-based variance reduction to SMC are described. 

\subsection{Sequential Monte Carlo}

SMC samplers are naturally adaptive and parallelizable alternatives to standard MCMC for sampling from the posterior of static Bayesian models \citep{DelMoral2006}. A set of $N$ weighted samples, $\{W_j^i,\vtheta_j^i\}_{i=1}^N$, are moved through a sequence of distributions, $p_{t_j}(\vtheta|\vect{y})$, for $j=0,\ldots,T$. The distributions $p_{t_j}$, henceforth $p_j$ for brevity, are properly normalized and $\eta_j$ represents the unnormalized distributions, i.e.\ $p_j\propto \eta_j$ for $j=0,\ldots,T$. The samples, or particles, are moved through these distributions using importance sampling, resampling and move steps.

The importance sampling step reweights particles $\{W_j^i,\vtheta_j^i\}_{i=1}^N$ from $p_{j-1}$ to target $p_j$ using
\begin{equation*}
w_j^i = W_{j-1}^i \frac{\eta_j(\vtheta_{j-1}^i)}{\eta_{j-1}(\vtheta_{j-1}^i)},
\end{equation*}
for $i=1,\ldots,N$, where $w_j^i$ is the unnormalized weight for particle $\vtheta_j^i = \vtheta_{j-1}^i$ and $W_0^i = 1/N$ if independent and identically distributed samples are drawn from the initial distribution $p_0$. The normalized weights are $W_j^i = w_j^i/\sum_{i=1}^N w_j^i$. Resampling, most commonly multinomial resampling, is used to remove particles with negligible weights and replicate particles with high weights. After resampling, the weights are set to $1/N$. Finally a move step, most commonly in the form of several iterations of a $p_j$-invariant MCMC kernel, is used to diversify the particles. Derivative based proposals have recently been used for the MCMC kernel in SMC \citep{Sim2012}, which means that ZV-CV can easily be performed on expectations with respect to $p_j$ for $j=1,\ldots,T$. 

It is straightforward to adapt SMC algorithms online. Recent work has proposed adapting the MCMC kernel parameters online using the population of particles. For example, \citet{Fearnhead2013} and \citet{Salomone2018} propose adaptation methods for generic MCMC kernels. \citet{Buchholz2018} propose methods for performing the notoriously challenging tuning of HMC kernel parameters in SMC. It is also possible to adaptively choose whether to perform the resampling and move steps based on some measure of the weight degeneracy \citep{DelMoral2012}. In this context, the weights are updated when resampling and move steps are not performed but the particle values remain the same. 

Commonly used sequences for $p_j$ in the literature are data annealing and likelihood annealing. In data annealing SMC \citep{Chopin2002}, the data are introduced sequentially so the targets are $p_j(\vtheta|\vect{y}_{1:j})$ where $\vect{y}_{1:j}$ denotes the first $j$ data points. Likelihood annealing smoothly introduces the effect of the likelihood to help explore complex targets \citep{Neal2001} through the sequence $p_j(\vtheta|\vect{y}) = \ell(\mathbf{y}|\vtheta)^{t_j} p_0(\vtheta)/Z_j$, the same sequence that is used in thermodynamic integration (TI, \citet{Gelman1998,Ogata1989}). Unlike in the MCMC setting, where $t_j$ for $j=0,\ldots,T$ needs to be fixed for TI, the inverse temperatures in SMC can easily be adapted online, for example using the methods described in \citet{Jasra2011} and detailed in Appendix \ref{app:Temperatures}. We use likelihood annealing SMC in all examples for these reasons.

\subsection{Handling Weights in ZV-CV}
SMC used weighted particle sets $\{W_i,\vtheta_i\}_{i=1}^N$. A weighted least squares can be performed in ZV-CV to take into account these weights. Recall that the optimization problem for unweighted samples $\{\vtheta_i\}_{i=1}^N$ is
\begin{equation*}
(\hat{c},\hat{g}) \in \argmin_{\substack{c \in \mathbb{R} \\ g \in \mathcal{G}}} \frac{1}{N} \sum_{i=1}^N \left[ \varphi(\bm{\theta}_i) - c + \mathcal{L}g(\bm{\theta}_i)\right]^2 + \lambda\text{PEN}(g).
\end{equation*}
With a weighted set of samples $\{W_i,\vtheta_i\}_{i=1}^N$, the optimization problem becomes 
\begin{equation}
(\hat{c},\hat{g}) \in \argmin_{\substack{c \in \mathbb{R} \\ g \in \mathcal{G}}} \frac{1}{N} \sum_{i=1}^N W_i \left[ \varphi(\bm{\theta}_i) - c + \mathcal{L}g(\bm{\theta}_i)\right]^2 + \lambda\text{PEN}(g).\label{eqn:weightedZVCV}
\end{equation}

\subsection{Posterior Expectations}

The most straightforward way to estimate posterior expectations in SMC is to perform Monte Carlo integration using the final set of particles obtained after all $T$ iterations. When resampling is performed at the final iteration, as we have done here, the final set of samples is the unweighted set $\{\vtheta_T^i\}_{i=1}^N$. The vanilla Monte Carlo estimator for the $j$th marginal posterior mean, $\mathbb{E}_p[\theta[j]]$, is $\widehat{\vtheta[j]}=\sum_{i=1}^N \vtheta_T^i[j]$, where $\vtheta[j]$ denotes the $j$th marginal. We apply ZV-CV and CF to improve this estimator.

It is also possible to estimate the posterior mean by using weighted particles from previous likelihood annealing targets. This recycling would extend the work of \citet{Briol2017} for improving tail coverage with the split estimator and could also improve the performance of higher order polynomials due to higher degrees of freedom. The weighted least squares approach in \eqref{eqn:weightedZVCV} could be applied in this setting.

\subsection{Evidence Estimation}\label{apps:Evidence}
An added benefit of SMC samplers over alternatives like standard MCMC is that an estimate of the normalising constant is produced as a by-product in SMC. Two evidence estimators that can be obtained as a by-product of SMC are described below. Both of these evidence estimators are based on a likelihood annealing schedule, but applications of Stein-based variance reduction techniques to the data annealing SMC evidence estimator for streaming data would also be possible.

\subsubsection{Controlled Thermodynamic Integration}

\citet{Oates2016} and \citet{Oates2017a} have applied ZV-CV and CF, respectively, to the power posterior estimator for the evidence \citep{Friel2008}. The estimator,
\begin{align}\label{eqn:TI_integral}
\log Z &= \int_0^1 \mathbb{E}_{p_t}[\log \ell(\mathbf{y}|\vtheta)]\mathrm{d}t,
\end{align}
is based on TI and it gives the log evidence as an integral with respect to the inverse temperature $t$, where $p_t=\ell(\mathbf{y}|\vtheta)^{t} p(\vtheta)/Z_t$. For points in a discrete set $\{t_j\}_{j=0}^T$ of inverse temperatures where $0= t_0<\ldots<t_T=1$, this integral is estimated using quadrature methods.

The second order quadrature method of \citet{Friel2014} estimates \eqref{eqn:TI_integral} using
\begin{align}\label{eqn:TI}
\widehat{\log Z} &= \sum_{j=0}^{T-1} \frac{t_{j+1}-t_j}{2} \left(\mathbb{E}_{p_t}[\log \ell(\mathbf{y}|\vtheta)] + \mathbb{E}_{p_{t+1}}[\log \ell(\mathbf{y}|\vtheta)]\right) \nonumber\\
& \phantom{{}= \sum_{j=0}^{T-1}}- \frac{(t_{j+1}-t_j)^2}{12}\left(\mathbb{V}_{p_{t+1}}[\log \ell(\mathbf{y}|\vtheta)] - \mathbb{V}_{p_t}[\log \ell(\mathbf{y}|\vtheta)] \right).
\end{align}
A simpler, first order quadrature approximation which is equivalent to the first sum in \eqref{eqn:TI} was described in \citet{Friel2008}.

It is straightforward to apply ZV-CV to \eqref{eqn:TI} by noticing that the estimator is simply a sum of expectations where
\begin{align*}
\mathbb{V}_{p_t}[\log \ell(\mathbf{y}|\vtheta)]=\mathbb{E}_{p_t}\left[(\log \ell(\mathbf{y}|\vtheta)-\mathbb{E}_{p_t}[\log \ell(\mathbf{y}|\vtheta)])^2\right]. 
\end{align*}
\citet{Oates2016} refer to the use of ZV-CV for this purpose as controlled TI (CTI) and they implement CTI in an MCMC framework. In the MCMC context, the inverse temperatures are fixed prior to the runs, making it difficult to balance low quadrature bias from using a large number of inverse temperatures with low computational effort from using a small number of inverse temperatures. Furthermore, sampling from the target distributions $p_{t_j}$ for $j=1,\ldots,T$ requires tuning, which is often done manually in MCMC. One contribution of our work is to use the CTI estimator in the SMC framework which allows for online choice of the inverse temperature schedule and online tuning of the MCMC proposal. Another contribution, which is described in Appendix \ref{app:Temperatures}, is a method for adjusting inverse temperatures after the MCMC or SMC runs have completed. We note that \eqref{eqn:TI_integral} has been used in the SMC framework by \citet{Zhou2012}, but CTI has not previously been used in SMC.

\subsubsection{SMC Evidence Estimator}

The standard SMC evidence estimator \citep{DelMoral2006} is based on the telescoping product $Z_T/Z_0=\prod_{j=1}^T Z_j/Z_{j-1}$. Assuming that independent and identically distributed samples are drawn from the prior $p_0$, so $Z_0=1$, the normalising constant can be written as the product of expectations as follows:
\begin{align}\label{eqn:zSMC}
Z=\prod_{j=1}^T\mathbb{E}_{p_{j-1}}\left[ \frac{\eta_j(\vtheta)}{\eta_{j-1}(\vtheta)} \right].
\end{align}
The standard estimator, $\widehat{Z} = \prod_{j=1}^T\sum_{i=1}^N W_{j-1}^i \ell(\vect{y}|\vtheta_{j-1}^i)^{t_j - t_{j-1}}$, is unbiased when adaptive methods are not used. We propose the use of ZV-CV on each of the expectations in \eqref{eqn:zSMC} to obtain a lower variance estimator, i.e.\ we take $p(\vtheta)$ in Section 2 of the main paper to be the power posterior $p_j(\vtheta|\vect{y})$. Although ZV-CV and CF can lead to unbiased estimators for each individual expectation, it is not clear whether the product of these estimators, $\hat{Z}$, remains unbiased. Nevertheless, we find in practice that the SMC estimator with ZV-CV has lower mean square error than the SMC estimator without control variates.

\subsubsection{Comparisons}

The CTI estimator appears to be more amenable to a control variates treatment, so we propose its use over the SMC estimator when practitioners wish to make the best use of derivatives for evidence estimation. The two evidence estimators both involve expectations of non-linear transformations to $\ell(\vect{y}|\vtheta)$, but the integrand is on the logarithmic scale for the CTI estimator which may make it a simpler function to estimate. However, the results for the challenging ordinary differential equation example in Appendix \ref{app:ODE} indicate that the SMC estimator may be preferable when the prior is highly diffuse. In this context, extreme values in the draws from the prior can lead to high variance in the CTI estimator.

When adaptive resampling is performed, the particles do not have equal weights at every iteration. ZV-CV is able to take into account these weights by using weighted means and weighted linear regression to estimate the ZV-CV coefficients $\vect{\beta}$. In the combined CF estimator, the information contained in the weights is lost due to its interpolating property. This may explain its poor performance for evidence estimation in Section 4.1 of the main paper.

\subsection{Post-hoc Temperature Choice}\label{app:Temperatures}

Using an insufficient number of inverse temperatures can lead to significant bias in the power posterior log evidence estimator. However, it is difficult to know \textit{a priori} how many inverse temperatures will be required to achieve reasonably small bias. This appendix briefly describes some existing methods for choosing the inverse temperature schedule before describing our post-hoc approach. The proposed method is useful for both CTI and SMC evidence estimation.

The simplest approach for choosing the inverse temperatures is to use a fixed schedule, for example $t_j = (j/T)^5$ for $j = 0,\ldots,T$ \citep{Friel2008}. If this schedule is conservative in that $T$ is very high, then some costly resampling and move steps can be avoided by performing these steps only when an approximation to the effective sample size (ESS) becomes low. The ESS is the number of independent samples from the target that would be required to achieve the same variance of the estimator and the ESS at inverse temperature $t_j$ is approximated in SMC by $1/\sum_{i=1}^N (W_j^i)^2$.

\citet{Friel2014} describe a method for adaptively choosing the inverse temperature schedule with the goal of minimising the discretization error in the power posterior log evidence estimator. Their method is used to calibrate the inverse temperatures prior to implementing the full sampler and it may be useful when evidence estimation is the primary focus. 

In the SMC context, the most popular method for adaptively choosing the inverse temperatures is based on fixing the approximated ESS at $\rho N$ using the bisection method, where $0 < \rho < N$ \citep{Jasra2011}. This approach maintains a fixed discrepancy between $p_{j-1}$ and $p_j$ for $j=1,\ldots,T$ when resampling and move steps are performed at each iteration. \citet{Zhou2015} use the conditional ESS (CESS),
\begin{align*}
\text{CESS}_{t_j} = \frac{N\left(\sum_{i=1}^N W_{j-1}^i \frac{\eta_j(\vtheta_{j-1}^i)}{\eta_{j-1}(\vtheta_{j-1}^i)}\right)^2}{\sum_{i=1}^N W_{j-1}^i \left(\frac{\eta_j(\vtheta_{j-1}^i)}{\eta_{j-1}(\vtheta_{j-1}^i)}\right)^2},
\end{align*}
instead of the ESS when resampling is not performed at every inverse temperature, because this is a more accurate measure of the discrepancy between $p_{j-1}$ and $p_j$ when $p_{j-1}$ is approximated with a weighted sample.

Using the approaches above, it is difficult to be confident that the quadrature bias will be sufficiently low without being overly conservative. Performing ZV-CV on the power posterior log evidence estimator requires $T$ regressions to be performed, that is one at each inverse temperature regardless of whether resample and move steps were performed. A conservative choice of inverse temperature schedule increases the post-processing time in ZV-CV.

We propose a post-hoc method to adjust the inverse temperatures when the original choice is either not conservative enough or too conservative. To start with, any approach can be used to provide $T$ distinct sets of particles and inverse temperatures,
\begin{equation}\label{eqn:original_temps}
\{      \{\vtheta_j^i\}_{i=1}^N,t_j   \}_{j=0}^T.
\end{equation}
Inverse temperatures at which the particles are not moved (for example due to adaptive resampling methods in SMC) are not included in these $T$ inverse temperatures. If \eqref{eqn:original_temps} is too conservative or not conservative enough, then a new set of inverse temperatures $\{\tilde{t}_{j}\}_{j=0}^{\tilde{T}}$ is selected as follows.

Given an inverse temperature $\tilde{t}_{j-1}$ (starting at $\tilde{t}_0=0$), the bisection method is used to select $\tilde{t}_j$ such that $\text{CESS}_{\tilde{t}_j} = \tilde{\rho} N$. This process continues until an inverse temperature of $\tilde{t}_{\tilde{T}} = 1$ satisfies $\text{CESS}_{\tilde{t}_{\tilde{T}}} \geq \tilde{\rho} N$. Each of the inverse temperatures must be assigned a relevant particle population and this is done by selecting population $\{\vtheta_k^i\}_{i=1}^N$ such that $t_k \leq \tilde{t}_j$. The new population is
\begin{equation}
\{      \{\vtheta_{\text{argmax}_k (t_k|\tilde{t_j}\geq t_k)}^i\}_{i=1}^N,\tilde{t}_j   \}_{j=0}^{\tilde{T}}.
\end{equation}
Put simply, the method involves choosing inverse temperatures post-hoc so that the CESS is fixed at $\tilde{\rho} N$ where $ 0 < \tilde{\rho} < N$. The inverse temperatures $\{ t_j \}_{j=0}^T$ need not appear in $\{ \tilde{t}_j \}_{j=0}^{\tilde{T}}$, but they may for some choices of initial inverse temperatures and $\tilde{\rho}$.

\subsection{Implementation Details} \label{app:Implementation}

The adaptive SMC methods used to select the tuning parameters and inverse temperatures are described in \citet{Salomone2018} and \citet{Jasra2011}, respectively. The post-hoc method for adapting the inverse temperatures is described in Appendix \ref{app:Temperatures}.

Table \ref{tab:Implementation} gives all tuning parameter specifications. The number of particles in the adaptive SMC run is $N$. Inverse temperatures in the adaptive SMC run are chosen to maintain an ESS of $\rho N$ and they are adjusted post-hoc to maintain a CESS of $\tilde{\rho} N$. The MCMC moves targeting $p_j$ use MALA proposals of the form
\begin{equation*}
q(\vtheta^*|\vtheta_j^i)=\mathcal{N}(\vtheta^{*};\vtheta_{j}^{i}+\frac{h_t^2}{2} \hat{\Sigma}_j\nabla_{\vtheta} \log{p_j(\vtheta|\vect{y})} ,h_j^2 \hat{\Sigma}_j),
\end{equation*}
where $\hat{\Sigma}_j$ is the empirical covariance and $h_j$ is a tuning parameter. We specify a set of 20 values which are log-uniform on the range of $h_{\text{min}}$ to $h_{\text{max}}$ and, following \citet{Salomone2018}, we select the value which maximizes the highest median estimated expected square jumping distance (ESJD, \citet{Pasarica2010}). Finally, we choose the number of MCMC repeats so that a given percentage of particles have a total absolute jumping distance greater than the mean Mahalanobis distance between particles before resampling \citep{Salomone2018}.

\begin{table}[!h]
\centering
\caption{Details of the adaptive SMC and post-hoc method tuning parameters for each example.}
\label{tab:Implementation}
\begin{tabular}{lrrrrrr}
  \hline
example  &  $N$  &   $h_{\text{min}}$   &   $h_{\text{max}}$   &   $\rho$   &   $\tilde{\rho}$  & $\% >$ median \\
  \hline
  Recapture & 10000 & 0.01 & 1 & 0.5 & 0.9 & 0.5 \\
  Cox & 10000 & 0.01 & 1 & 0.5 & 0.9 & 0.5 \\
  Van Der Pol & 1000 & 0.01 & 2 & 0.9 & 0.99 & 0.5 \\
  Sonar & 10000 & 0.01 & 1 & 0.5 & 0.9 & 0.5 \\
  ODE & 1000 & 0.01 & 1 & 0.5 & 0.9 & 0.5 \\
   \hline
\end{tabular}
\end{table}

To avoid confounding the effects of ZV-CV with the effects of kernel parameter and inverse temperature choice, we do a single adaptive SMC run for each example. The inverse temperatures and kernel parameters from this run are then used in 100 independent SMC runs for each value of $N$. 

The full set of inverse temperatures are used for both the CTI and SMC estimators. Additional inverse temperatures without resample and move steps do not improve the vanilla SMC evidence estimator, but we find that they can lead to substantial reductions in the ZV-CV SMC evidence estimator. To improve stability, the ZV-CV regression for the SMC evidence estimator is performed using the integrand divided by its maximum value and the results are adjusted to correct for this. 

\section{Additional Results for the Recapture Example}\label{app:recapture}

\begin{figure}[h!]
\centering
\includegraphics[width=1\textwidth]{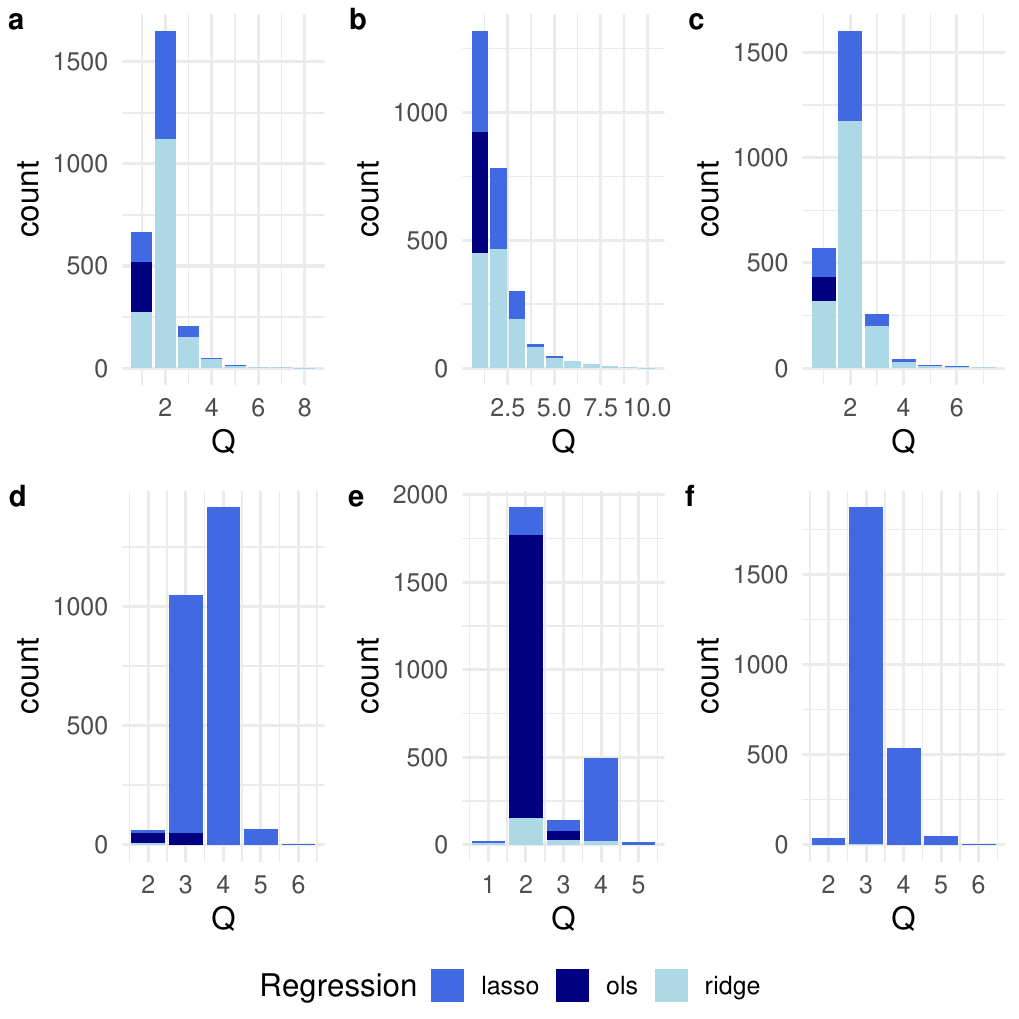}
\caption{Recapture example: The selected control variates based on cross-validation for (a) The $\mathbb{E}$ terms in CTI \eqref{eqn:TI} with $N=50$, (b) The $\mathbb{V}$ terms in CTI \eqref{eqn:TI} with $N=50$, (c) The $\mathbb{E}$ terms in \eqref{eqn:zSMC} with $N=50$, (d)-(f) the same for $N=1000$.}
\label{fig:recapture_evidence_selected}
\end{figure}

The selected control variates for the 1100 posterior expectations at $N=50$ and $N=1000$ can be seen in Table \ref{tab:recapture_p_selection}. Control variate selection for evidence estimatation at $N=50$ and $N=1000$ is shown in Figure \ref{fig:recapture_evidence_selected}. It is clear that higher order polynomials are selected for larger $N$, and lasso is selected more often than ridge regression. These general trends are consistent with the best performing fixed methods.

\begin{table}
\centering
\caption{Recapture example: control variate selection for marginal posterior expectations using cross-validation. Each value of $N$ has a total of 1100 expectations, that is 100 independent runs are used to estimate each marginal posterior expectation.}
\label{tab:recapture_p_selection}
\begin{subtable}{.45\textwidth}
\centering
\caption{$N=50$}
\begin{tabular}{rrrrr}
  \hline
 & $\text{sub}_1\text{-ZV}$ & ZV & $l\text{-ZV}$ & $r\text{-ZV}$ \\ 
  \hline
1 &  1 &  108 & 582 &   71 \\ 
  2 &  4 &   0 & 301 &   5 \\ 
  3 &   3 &   0 & 19 &   0 \\ 
  4 &   2 &   0 &  0 &   0 \\ 
  5 &   3 &   0  &  0 &   0 \\ 
  6 &   1 &   0  &  0 &   0 \\ 
   \hline
\end{tabular}
\end{subtable}
\begin{subtable}{.45\textwidth}
\centering
\caption{$N=1000$}
 \begin{tabular}{rrrr}
  \hline
$Q$ & ZV & $l\text{-ZV}$ & $r\text{-ZV}$ \\ 
  \hline
2 &   0 & 24 &     0 \\ 
  3 & 632 & 307 &  2 \\ 
  4 &   0  & 80 &   0 \\ 
  5 &   0 &  43  &   0 \\ 
  6 &   0 &  12  &   0 \\
   \hline
\end{tabular}
\end{subtable}
\end{table}

\clearpage

\section{Additional Results for the Log-Gaussian Cox Process Example}\label{app:LogCoxApp}

A gold standard estimate is required for all expectations of interest in order to assess the statistical efficiency of the different methods. For the Log-Gaussian Cox Process example, the gold standard of approximation comes from unbiased Riemannian-manifold Hamiltonian Monte Carlo (RM-HMC) with coupling as per \citep{Heng2019}. Control variates based on minimisation of an upper bound on the variance are also used, as described in \citet{South2019discussion}. The resulting estimators are unbiased, meaning that many cores can be used in parallel to estimate the expectations. 

For all three models, the estimates were averaged over 30 repeats where the following settings were used. The RM-HMC part of the algorithm had a step size of 0.11 with 10 steps. The probability of selecting the coupled random walk Metropolis Hastings (RWMH) algorithm rather than the RM-HMC sampler was 0.05 and the proposal standard deviation of the RWMH was 0.001. All algorithms had a burn-in of 70 iterations. The total number of MCMC iterations for $d=16$, $d=64$ and $d=256$ were $100,000$, $15,000$ and $4000$, respectively. Each of the 30 repeats involved running $R=32$ independent sets of coupled Markov chains. Control variates were applied such that half of the $R=32$ lots of samples were used to estimate control variate coefficients for the other half of the samples. This was done to maintain unbiasedness. The specific classes of control variates used were third order ZV-CV, second order ZV-CV and first order ZV-CV for $d=16$, $d=64$ and $d=256$, respectively.

Tables \ref{tab:cox16_1000}-\ref{tab:cox256_10000} show the mean relative statistical, computational and overall efficiency for posterior expectations in dimensions $d=16$, $d=64$ and $d=256$ when $N=1,000$ and $N=10,000$. For $N=1,000$, the overall efficiency of the best novel method compared to the best existing method (vanilla MC, ZV-CV with OLS or CF) is a factor of over 600 for $d=16$, over 600 for $d=64$ and over 1.35 for $d=256$. For $N=10,000$, the overall efficiency of the best novel method compared to the best existing method is a factor of over 3 for $d=16$, over 4 for $d=64$ and over 0.98 (a slight decrease) for $d=256$.


\begin{table*}
\centering
\caption{16-dimensional Cox example: efficiency for marginal posterior expectations when $N=1,000$, averaged over results for all 16 parameters. A ``$-$''' indicates that the population size $N=1,000$ is insufficient for standard ZV-CV. We refer the reader to the beginning of Section 4 for acronym definitions.} 
\label{tab:cox16_1000}
\begin{tabular}{llrrr}
Approach & $Q$ & Stat. Efficiency & Comp. Efficiency & Overall Efficiency\\ \hline
Vanilla & NA & $1.0 \times 10^0$ & $1.0000$ & $1.0 \times 10^0$\\
CF & NA & $7.8 \times 10^4$ & $0.6470$ & $5.0 \times 10^4$\\ \hline
\zv{1} & 1 & $2.3 \times 10^1$ & $0.9999$ & $2.3 \times 10^1$\\
\zvl{1} & 1 & $2.3 \times 10^1$ & $0.9960$ & $2.3 \times 10^1$\\
\zvr{1} & 1 & $2.3 \times 10^1$ & $0.9959$ & $2.3 \times 10^1$\\ \hline
\zv{2} & 2 & $4.5 \times 10^3$ & $0.9991$ & $4.5 \times 10^3$\\
\zvl{2} & 2 & $4.1 \times 10^3$ & $0.9845$ & $4.1 \times 10^3$\\
\zvr{2} & 2 & $3.2 \times 10^3$ & $0.9271$ & $3.0 \times 10^3$\\ \hline
\zv{3} & 3 & $1.2 \times 10^5$ & $0.9847$ & $1.1 \times 10^5$\\
\zvl{3} & 3 & $1.9 \times 10^3$ & $0.8033$ & $1.6 \times 10^3$\\
\zvr{3} & 3 & $1.3 \times 10^3$ & $0.3542$ & $4.7 \times 10^2$\\ \hline
\zv{4} & 4 & $-$ & $-$ & $-$\\
\zvl{4} & 4 & $2.3 \times 10^3$ & $0.3965$ & $9.8 \times 10^2$\\
\zvr{4} & 4 & $7.5 \times 10^2$ & $0.1458$ & $1.1 \times 10^2$\\ \hline
\zvs{1}{1} & 1 & $2.3 \times 10^1$ & $0.9999$ & $2.3 \times 10^1$\\
\zvs{2}{1} & 2 & $5.4 \times 10^3$ & $0.9999$ & $5.4 \times 10^3$\\
\zvs{3}{1} & 3 & $3.8 \times 10^6$ & $0.9999$ & $3.8 \times 10^6$\\
\zvs{4}{1} & 4 & $6.8 \times 10^7$ & $0.9999$ & $6.8 \times 10^7$\\ \hline
crossval & NA & $6.9 \times 10^8$ & $0.0986$ & $6.2 \times 10^7$\\
\end{tabular}
\normalsize
\end{table*}

\begin{table*}
\centering
\caption{64-dimensional Cox example: efficiency for marginal posterior expectations when $N=1,000$, averaged over results for all 64 parameters. A ``$-$''' indicates that the population size $N=1,000$ is insufficient for standard ZV-CV. We refer the reader to the beginning of Section 4 for acronym definitions.} 
\label{tab:cox64_1000}
\begin{tabular}{llrrr}
Approach & $Q$ & Stat. Efficiency & Comp. Efficiency & Overall Efficiency\\ \hline
Vanilla & NA & $1.0 \times 10^0$ & $1.0000$ & $1.0 \times 10^0$\\
CF & NA & $2.2 \times 10^1$ & $0.9252$ & $2.1 \times 10^1$\\ \hline
\zv{1} & 1 & $1.4 \times 10^1$ & $0.9991$ & $1.4 \times 10^1$\\
\zvl{1} & 1 & $1.5 \times 10^1$ & $0.9980$ & $1.5 \times 10^1$\\
\zvr{1} & 1 & $1.4 \times 10^1$ & $0.9979$ & $1.4 \times 10^1$\\ \hline
\zv{2} & 2 & $-$ & $-$ & $-$\\
\zvl{2} & 2 & $2.4 \times 10^3$ & $0.9468$ & $2.3 \times 10^3$\\
\zvr{2} & 2 & $1.4 \times 10^1$ & $0.8604$ & $1.2 \times 10^1$\\ \hline
\zvs{1}{1} & 1 & $1.5 \times 10^1$ & $1.0000$ & $1.5 \times 10^1$\\
\zvs{2}{1} & 2 & $2.6 \times 10^3$ & $1.0000$ & $2.7 \times 10^3$\\
\zvs{3}{1} & 3 & $8.9 \times 10^3$ & $1.0000$ & $8.9 \times 10^3$\\
\zvs{4}{1} & 4 & $1.3 \times 10^4$ & $1.0000$ & $1.3 \times 10^4$\\ \hline
crossval & NA & $1.2 \times 10^4$ & $0.4684$ & $5.4 \times 10^3$\\
\end{tabular}
\normalsize
\end{table*}

\begin{table*}
\centering
\caption{256-dimensional Cox example: efficiency for marginal posterior expectations when $N=1,000$, averaged over results for all 256 parameters. We refer the reader to the beginning of Section 4 for acronym definitions.} 
\label{tab:cox256_1000}
\begin{tabular}{llrrr}
Approach & $Q$ & Stat. Efficiency & Comp. Efficiency & Overall Efficiency\\ \hline
Vanilla & NA & $1.0$ & $1.0000$ & $1.0$\\
CF & NA & $37.8$ & $0.9858$ & $37.3$\\ \hline
\zv{1} & 1 & $43.1$ & $0.9999$ & $43.1$\\
\zvl{1} & 1 & $59.7$ & $0.9995$ & $59.7$\\
\zvr{1} & 1 & $40.6$ & $0.9994$ & $40.6$\\ \hline
\zvs{1}{1} & 1 & $25.6$ & $1.0000$ & $25.6$\\
\zvs{2}{1} & 2 & $52.6$ & $1.0000$ & $52.6$\\
\zvs{3}{1} & 3 & $53.7$ & $1.0000$ & $53.7$\\
\zvs{4}{1} & 4 & $53.8$ & $1.0000$ & $53.8$\\ \hline
crossval & NA & $48.2$ & $0.9934$ & $47.9$\\
\end{tabular}
\normalsize
\end{table*}


\begin{table*}
\centering
\caption{16-dimensional Cox example: efficiency for marginal posterior expectations when $N=10,000$, averaged over results for all 16 parameters. We refer the reader to the beginning of Section 4 for acronym definitions.} 
\label{tab:cox16_10000}
\begin{tabular}{llrrr}
Approach & $Q$ & Stat. Efficiency & Comp. Efficiency & Overall Efficiency\\ \hline
Vanilla & NA & $1.0 \times 10^0$ & $1.0000$ & $1.0 \times 10^0$\\
CF & NA & $2.1 \times 10^7$ & $0.0171$ & $3.6 \times 10^5$\\ \hline
\zv{1} & 1 & $2.5 \times 10^1$ & $0.9999$ & $2.5 \times 10^1$\\
\zvl{1} & 1 & $2.5 \times 10^1$ & $0.9990$ & $2.5 \times 10^1$\\
\zvr{1} & 1 & $2.5 \times 10^1$ & $0.9990$ & $2.5 \times 10^1$\\ \hline
\zv{2} & 2 & $5.5 \times 10^3$ & $0.9992$ & $5.5 \times 10^3$\\
\zvl{2} & 2 & $4.9 \times 10^3$ & $0.9905$ & $4.8 \times 10^3$\\
\zvr{2} & 2 & $4.4 \times 10^3$ & $0.9841$ & $4.4 \times 10^3$\\ \hline
\zv{3} & 3 & $4.1 \times 10^6$ & $0.9760$ & $4.0 \times 10^6$\\
\zvl{3} & 3 & $3.0 \times 10^3$ & $0.8219$ & $2.5 \times 10^3$\\
\zvr{3} & 3 & $2.4 \times 10^3$ & $0.3579$ & $8.1 \times 10^2$\\ \hline
\zv{4} & 4 & $3.4 \times 10^7$ & $0.6730$ & $2.3 \times 10^7$\\
\zvl{4} & 4 & $6.3 \times 10^3$ & $0.3516$ & $2.4 \times 10^3$\\
\zvr{4} & 4 & $2.4 \times 10^3$ & $0.1221$ & $2.9 \times 10^2$\\ \hline
\zvs{1}{1} & 1 & $2.5 \times 10^1$ & $1.0000$ & $2.5 \times 10^1$\\
\zvs{2}{1} & 2 & $5.6 \times 10^3$ & $1.0000$ & $5.6 \times 10^3$\\
\zvs{3}{1} & 3 & $5.2 \times 10^6$ & $1.0000$ & $5.2 \times 10^6$\\
\zvs{4}{1} & 4 & $7.8 \times 10^7$ & $1.0000$ & $7.8 \times 10^7$\\ \hline
crossval & NA & $7.3 \times 10^8$ & $0.0774$ & $5.7 \times 10^7$\\
\end{tabular}
\normalsize
\end{table*}

\begin{table*}
\centering
\caption{64-dimensional Cox example: efficiency for marginal posterior expectations when $N=10,000$, averaged over results for all 64 parameters We refer the reader to the beginning of Section 4 for acronym definitions.} 
\label{tab:cox64_10000}
\begin{tabular}{llrrr}
Approach & $Q$ & Stat. Efficiency & Comp. Efficiency & Overall Efficiency\\ \hline
Vanilla & NA & $1.0 \times 10^0$ & $1.0000$ & $1.0 \times 10^0$\\
CF & NA & $2.6 \times 10^3$ & $0.3406$ & $8.8 \times 10^2$\\ \hline
\zv{1} & 1 & $1.6 \times 10^1$ & $0.9998$ & $1.6 \times 10^1$\\
\zvl{1} & 1 & $1.6 \times 10^1$ & $0.9993$ & $1.6 \times 10^1$\\
\zvr{1} & 1 & $1.6 \times 10^1$ & $0.9993$ & $1.6 \times 10^1$\\ \hline
\zv{2} & 2 & $3.1 \times 10^3$ & $0.9802$ & $3.0 \times 10^3$\\
\zvl{2} & 2 & $3.7 \times 10^3$ & $0.9373$ & $3.5 \times 10^3$\\
\zvr{2} & 2 & $2.7 \times 10^3$ & $0.3427$ & $8.3 \times 10^2$\\ \hline
\zvs{1}{1} & 1 & $1.6 \times 10^1$ & $1.0000$ & $1.6 \times 10^1$\\
\zvs{2}{1} & 2 & $2.8 \times 10^3$ & $0.9999$ & $2.8 \times 10^3$\\
\zvs{3}{1} & 3 & $8.7 \times 10^3$ & $0.9999$ & $8.7 \times 10^3$\\
\zvs{4}{1} & 4 & $1.3 \times 10^4$ & $0.9999$ & $1.3 \times 10^4$\\ \hline
crossval & NA & $1.3 \times 10^4$ & $0.1116$ & $1.2 \times 10^3$\\
\end{tabular}
\normalsize
\end{table*}

\begin{table*}
\centering
\caption{256-dimensional Cox example: efficiency for marginal posterior expectations when $N=10,000$, averaged over results for all 256 parameters. We refer the reader to the beginning of Section 4 for acronym definitions.} 
\label{tab:cox256_10000}
\begin{tabular}{llrrr}
Approach & $Q$ & Stat. Efficiency & Comp. Efficiency & Overall Efficiency\\ \hline
Vanilla & NA & $1.0$ & $1.0000$ & $1.0$\\
CF & NA & $150.5$ & $0.3847$ & $57.9$\\ \hline
\zv{1} & 1 & $104.6$ & $1.0000$ & $104.6$\\
\zvl{1} & 1 & $92.8$ & $0.9996$ & $92.8$\\
\zvr{1} & 1 & $103.0$ & $0.9996$ & $103.0$\\ \hline
\zvs{1}{1} & 1 & $27.6$ & $1.0000$ & $27.6$\\
\zvs{2}{1} & 2 & $51.0$ & $1.0000$ & $51.0$\\
\zvs{3}{1} & 3 & $51.7$ & $1.0000$ & $51.7$\\
\zvs{4}{1} & 4 & $51.7$ & $1.0000$ & $51.7$\\ \hline
crossval & NA & $49.2$ & $0.9964$ & $49.1$\\
\end{tabular}
\normalsize
\end{table*}

\clearpage

\section{Sonar Example}\label{app:Sonar}
Using all derivative information to perform ZV-CV with higher order polynomials may simply be unrealistic for some examples, due to storage constraints or to the number of regression parameters required. The 61-dimensional example below falls into this class of problems. A polynomial with $Q=2$ for this example requires a restrictive 1953 regression parameters while $Q=3$ and $Q=4$ polynomials require a potentially unrealistic number of samples with 41,664 and 677,040 regression parameters, respectively.

This binary regression problem is based on discriminating between sonar signals bouncing off a metal cylinder versus sonar signals bouncing off a roughly cylindrical rock. The 60 covariates represent the total energy within a given frequency band over a fixed period of time, with increasing aspect angle from covariate 1 to covariate 60. Observations $y_i$ for $i=1,\ldots,208$ are coded as 0 for rock and 1 for metal. The corresponding log likelihood is
\begin{equation}\label{eqn:logistic}
\log \ell(\vect{y},\vect{X}|\vtheta) = \sum_{i=1}^{208} \left(y_i \vect{X}_{i,\cdot}\vtheta - \log (1+\exp(\vect{X}_{i,\cdot}\vtheta) )\right),
\end{equation}
where $\vect{X} \in \mathbb{R}^{208 \times 61}$ is the matrix of covariates starting with a column of 1's for the intercept, $\vect{y} \in \mathbb{R}^{208}$ is the vector of indicators for the response and $\vtheta \in \mathbb{R}^{61}$ is the vector of coefficients. Using \eqref{eqn:logistic} with $\{0,1\}$ encoding of the response is equivalent to using $\log \ell(\vect{y},\vect{X}|\vtheta) = -\sum_{i=1}^{208} \log (1+\exp(-y_i\vect{X}_{i,\cdot}\vtheta ))$ with $\{-1,1\}$ encoding  (see e.g.\ \citet{Hastie2015}). Following \citet{Chopin2017}, we standardize the predictors (columns 2-61 of $\vect{X}$) to have standard deviation 0.5 and we use $\mathcal{N}(0,5^2)$ priors. The intercept has a $\mathcal{N}(0,20^2)$ prior. This prior specification is chosen over the Cauchy priors of \citet{Gelman2008} so that the expectations exist and boundary condition (2.4) from the main paper is satisfied. The data used here is from the UCI machine learning repository \citep{UCI} and was originally collected by \citet{Gorman1988}.

This example is more challenging than most standard logistic regression problems due to the the high number of covariates and high correlations between covariates \citep{Chopin2017}. The aspect angle is increasing from covariate 1 to covariate 60, so it is reasonable to assume that coefficients of nearby covariates will be correlated. This gives useful information in choosing the subset of parameters for \textit{a priori} ZV-CV. We perform ZV-CV with subsets of 1 or 5 parameters in estimating marginal posterior expectations. The polynomials with a subset of five parameters are based on the four closest angles, for example when estimating the posterior expectation of $\theta[30]$, the polynomial is a function of a subset of the five parameters $S = \{28,29,30,31,32\}$. The gold standard of posterior expectation in this example is the average posterior expectation across 100 independent SMC runs using $N=10000$ and ZV-CV with $Q=2$. Due to memory and time constraints, a maximum polynomial order of $Q=2$ was considered in the cross-validation.

Table \ref{tab:sonar} shows the mean statistical efficiency for posterior expectations. For small $N$, one can obtain better results using a subset of parameters than with \zvl{1}, whereas for large $N$ polynomials with $Q=2$ become more efficient. Out of the 6100 expectations for $N=50$, 83.9\% of the control variates selected based on cross-validation use the subset of five parameters, 3.1\% use the full 61 parameters with penalized regression methods and 13.1\% use the subset of one parameter. At $N=5000$, 97.5\% of selected control variates are \zvl{2}, 2.5\% are \zvr{2} and less than 1\% use $Q=1$. As suspected, due to the high dimension in this example, ZV-CV methods outperform CF.

\begin{table*}
\centering
\caption{Sonar example: statistical efficiency for marginal expectations, averaged over results for all 61 parameters. Results for individual parameters are similar. Blank values indicate that the number of samples is not sufficient for this order polynomial. Bold values indicate the most efficient control variate for a fixed $N$.} 
\label{tab:sonar}
\large
\resizebox{1\columnwidth}{!}{%
\begin{tabular}{rrrrrrrrrrr}
  \hline
$N$ & \zv{1} & \zvl{1} & \zvr{1} & \zv{2} & \zvl{2} & \zvr{2} & \zvs{1}{1} & \zvs{1}{5} & crossval & CF\\  
  \hline
50 &  & 1.3 & 1.1 &  & 1.1 & 1.1 & 1.2 & \textbf{1.9} & 1.7 & 1.0 \\ 
  100 & 4.3 & \textbf{4.5} & 4.2 &  & 1.1 & 1.1 & 1.1 & 2.1 & 2.0 & 1.2 \\ 
  500 & \textbf{11} & 10 & 10 &  & 2.9 & 2.0 & 1.2 & 2.2 & 11 & 7.8 \\ 
  5000 & 13 & 13 & 13 & \textbf{45} & 40 & 43 & 1.2 & 2.2 & 40 & 35 \\ 
   \hline
\end{tabular}}
\normalsize
\end{table*}

Table \ref{tab:sonar_overall} shows the overall efficiency for each posterior expectation. Like the recapture example, the likelihood function in this example is relatively inexpensive. The proposed methods offer up to two orders of magnitude in overall efficiency compared to vanilla MC, and \textit{a priori} ZV-CV is the best performing method for $N=50$.

\begin{table*}
\centering
\caption{Sonar example: overall efficiency for marginal expectations, averaged over results for all 61 parameters. Blank values indicate that the number of samples is not sufficient for this order polynomial. Bold values indicate the most efficient control variate in terms of overall efficiency for a fixed $N$.} 
\label{tab:sonar_overall}
\begin{tabular}{rrrrrrrrrrr}
  \hline
$N$ & \zv{1} & \zvl{1} & \zvr{1} & \zv{2} & \zvl{2} & \zvr{2} & \zvs{1}{1} & \zvs{1}{5} & crossval & CF\\ 
  \hline
50 &  & 1.3 & 1.1 &  & 1.0 & 1.0 & 1.2 & \textbf{1.9} & 1.4  & 1.0\\ 
  100 & 4.3 & \textbf{4.5} & 4.2 &  & 1.1 & 1.1 & 1.1 & 2.1 & 1.8 & 1.2 \\ 
  500 & \textbf{11} & 10 & 10 &  & 2.8 & 2.0 & 1.2 & 2.2 & 9.9 & 7.6 \\ 
  5000 & 13 & 13 & 13 & \textbf{44} & 34 & 32 & 1.2 & 2.2 & 26 & 20 \\
   \hline
\end{tabular}
\end{table*}

\clearpage

\section{Van der Pol Example}\label{ssec:VanDerPol}
This example based on the Van der Pol oscillatory differential equations \citep{VanDerPol1926} demonstrates the potential to improve the performance of ZV-CV by using higher order polynomials. Regularization is not required in this one-dimensional example.

In the Van der Pol oscillatory differential equations, position $x(t)$ is governed by the second order differential equation $\frac{d^2x}{dt^2} - \theta(1-x^2)\frac{dx}{dt} + x = 0$, which can also be written as a system of first order differential equations,
\begin{align*}
\frac{dx_1}{dt}&=x_2\\
\frac{dx_2}{dt}&=\theta(1-x_1^2)x_2 - x_1,
\end{align*}
where $x_1=x$ and $x_2 = dx/dt$. The parameter $\theta$ represents the non-linearity of the system and the damping strength. We use the same data as \citet{Oates2017a} which is based on $\theta =1$,  initial position $x_1=0$, initial velocity $x_2=2$ and noisy observations $y_t\sim\mathcal{N}(x_1(t),0.1^2)$ recorded at times $t=0,1,\ldots,10$. The prior is $\log(\theta) \sim \mathcal{N}(0,0.25^2)$. Derivatives are obtained by augmenting the system of differential equations with sensitivity equations (see e.g.\ Appendix 7 of \citet{Oates2017a}).

\citet{Oates2017a} found that second order polynomials resulted in mediocre performance for evidence estimation in this example. \citet{Oates2017a} used population MCMC with within-temperature and between-temperature proposals to obtain samples from the power posteriors. The sampling algorithm used here is SMC, which is simpler and easier to adapt. Given that the prior is such that $\theta>0$, we have performed a log transform so that the MCMC proposals and ZV-CV are both based on $\psi = \log\theta$. Expectations are calculated using (2.5) of the main paper, for example the posterior expectation of $\theta$ can be expressed as $\mathbb{E}_{p_\psi}[\exp(\psi)]$.

The gold standard evidence estimate\footnote{This evidence estimate differs from the estimates in \citet{Oates2017a}. The code available at \url{https://www.imperial.ac.uk/inference-group/projects/monte-carlo-methods/control-functionals/} is missing a square root in the normal probability density function which appears in the likelihood function.} for this example is based on numerical integration. For the posterior mean, the gold standard of $\widehat{\theta}$ is based on the mean of 100 estimates using \zv{9} at $N=2000$.

\subsubsection{Posterior Expectations}

\begin{figure}
\centering
\subcaptionbox{Statistical Efficiency\label{subfig:vdp_p_eff}}{\includegraphics[width=0.3\textwidth]{VanDerPol_p_stat}}
\subcaptionbox{Polynomial Order Selection\label{subfig:vdp_p_cross}}{\includegraphics[width=0.3\textwidth]{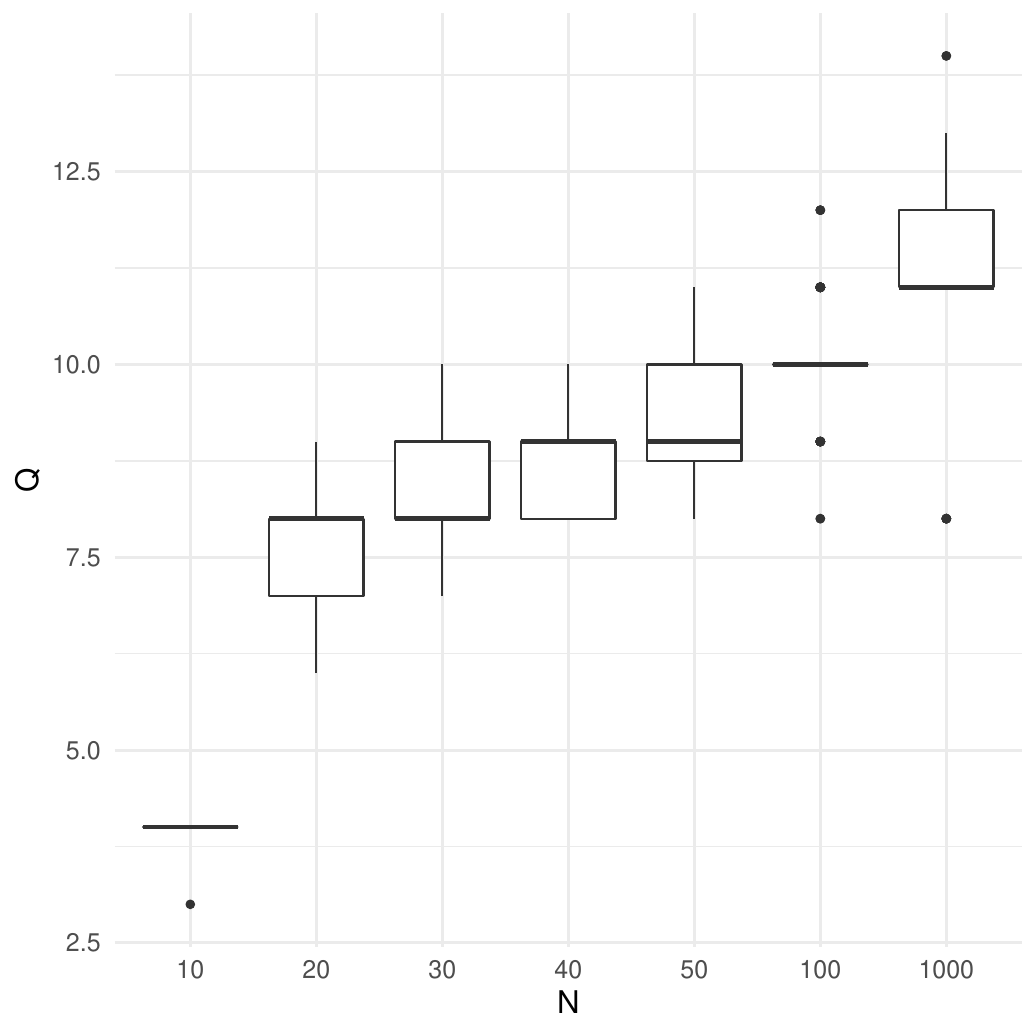}}
\caption{Van der Pol example: (a) statistical efficiency for $\widehat{\mathbb{E}_p[\theta]}$ and (b) the polynomial orders selected using cross-validation for different $N$.}
\label{fig:VanDerPol_posterior}
\end{figure}

With this simple integrand, ZV-CV is able to perform extremely well and outperforms CF. It is apparent from Figure \ref{fig:VanDerPol_posterior} that larger $Q$ is preferable for this problem, with the selected polynomial order from cross-validation increasing with $N$. Here the likelihood function is expensive so the majority of the time is taken up by sampling and the difference between the two measures of efficiency is negligible.

\begin{figure}
\centering
\subcaptionbox{Statistical Efficiency\label{subfig:vdp_p_eff}}{\includegraphics[width=0.3\textwidth]{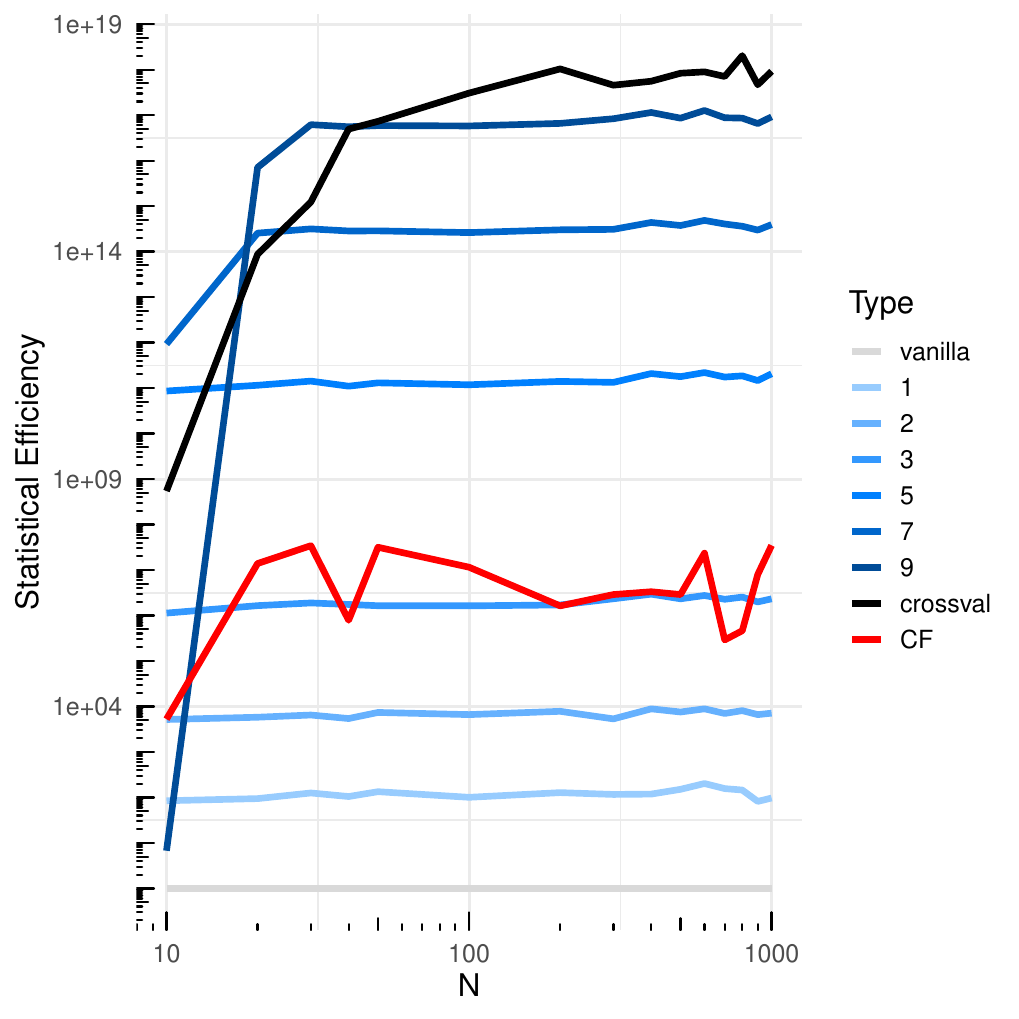}}
\subcaptionbox{Overall Efficiency\label{subfig:vdp_p_eff}}{\includegraphics[width=0.3\textwidth]{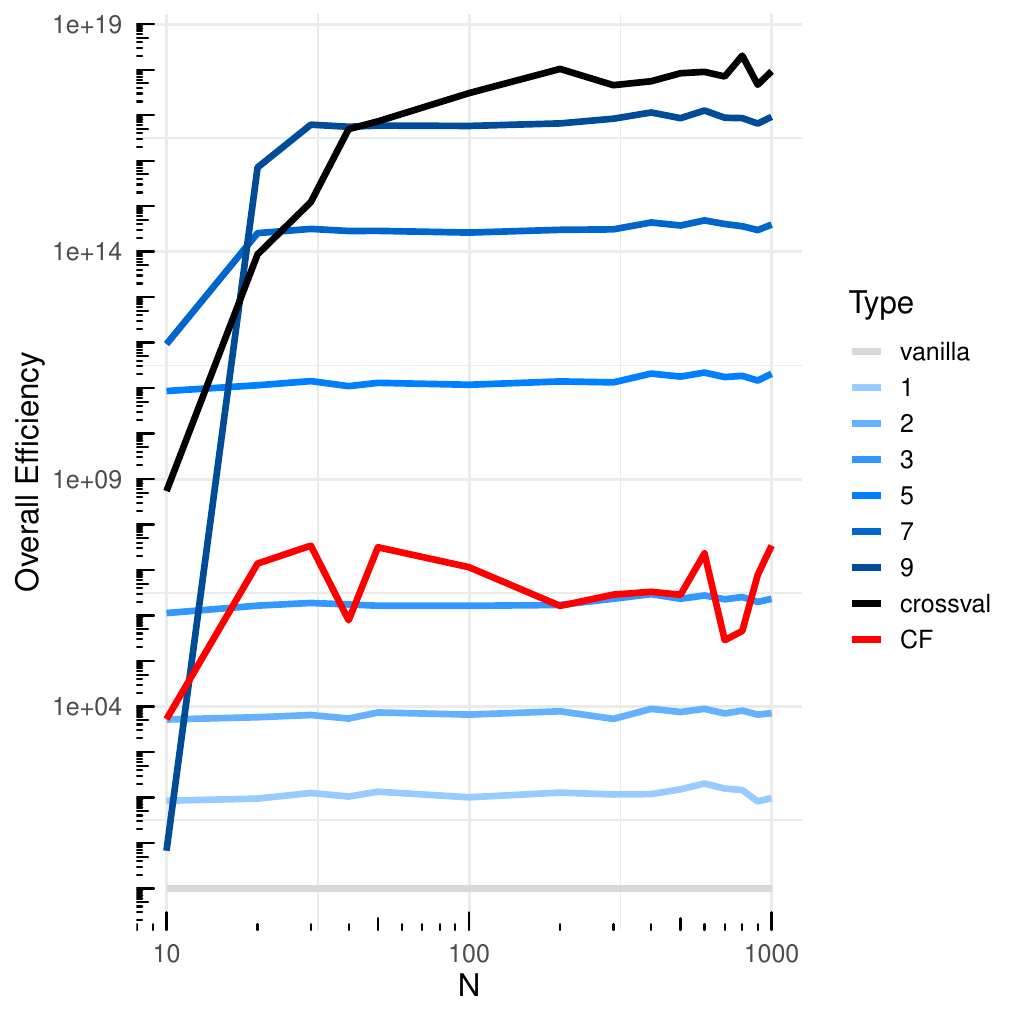}}
\caption{Van der Pol example: (a) statistical efficiency for $\widehat{\mathbb{E}_p[\theta]}$ and (b) overall efficiency for $\widehat{\mathbb{E}_p[\theta]}$.}
\label{fig:VanDerPol_posterior_overall}
\end{figure}

\subsubsection{Evidence Estimation}

\begin{figure*}
\centering
\subcaptionbox{Statistical efficiency CTI\label{subfig:vdp_CTI_eff}}{\includegraphics[width=0.33\textwidth]{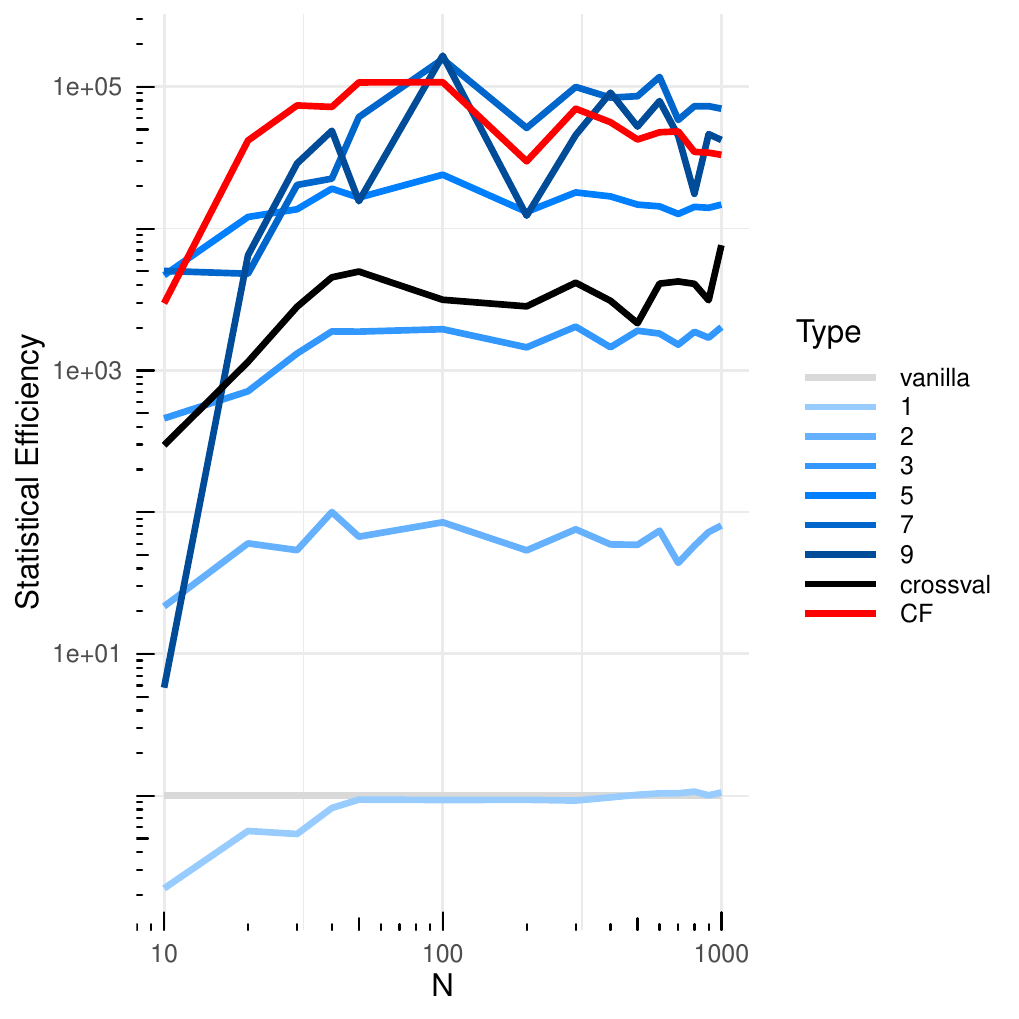}}
\subcaptionbox{Overall efficiency CTI\label{subfig:vdp_CTI_eff}}{\includegraphics[width=0.33\textwidth]{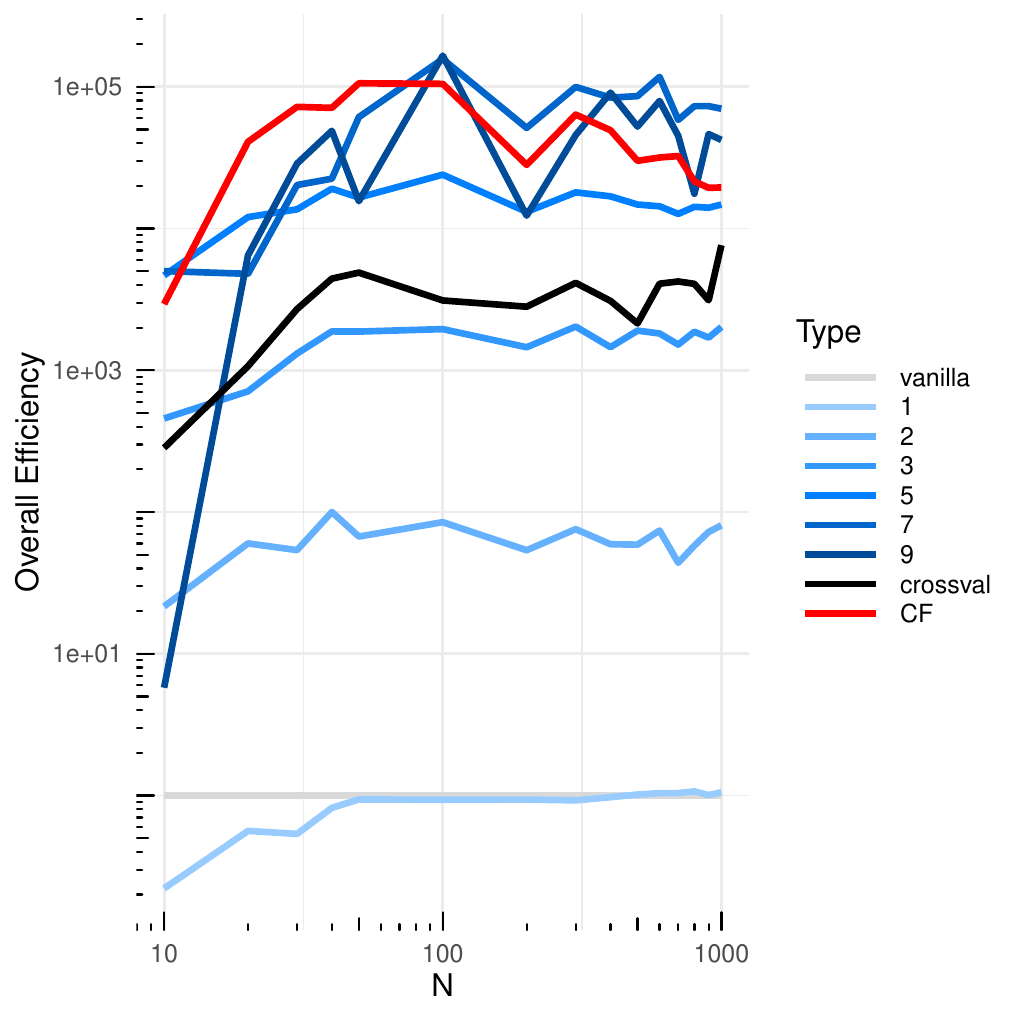}}\\
\subcaptionbox{Statistical efficiency SMC\label{subfig:vdp_SMC_eff}}{\includegraphics[width=0.33\textwidth]{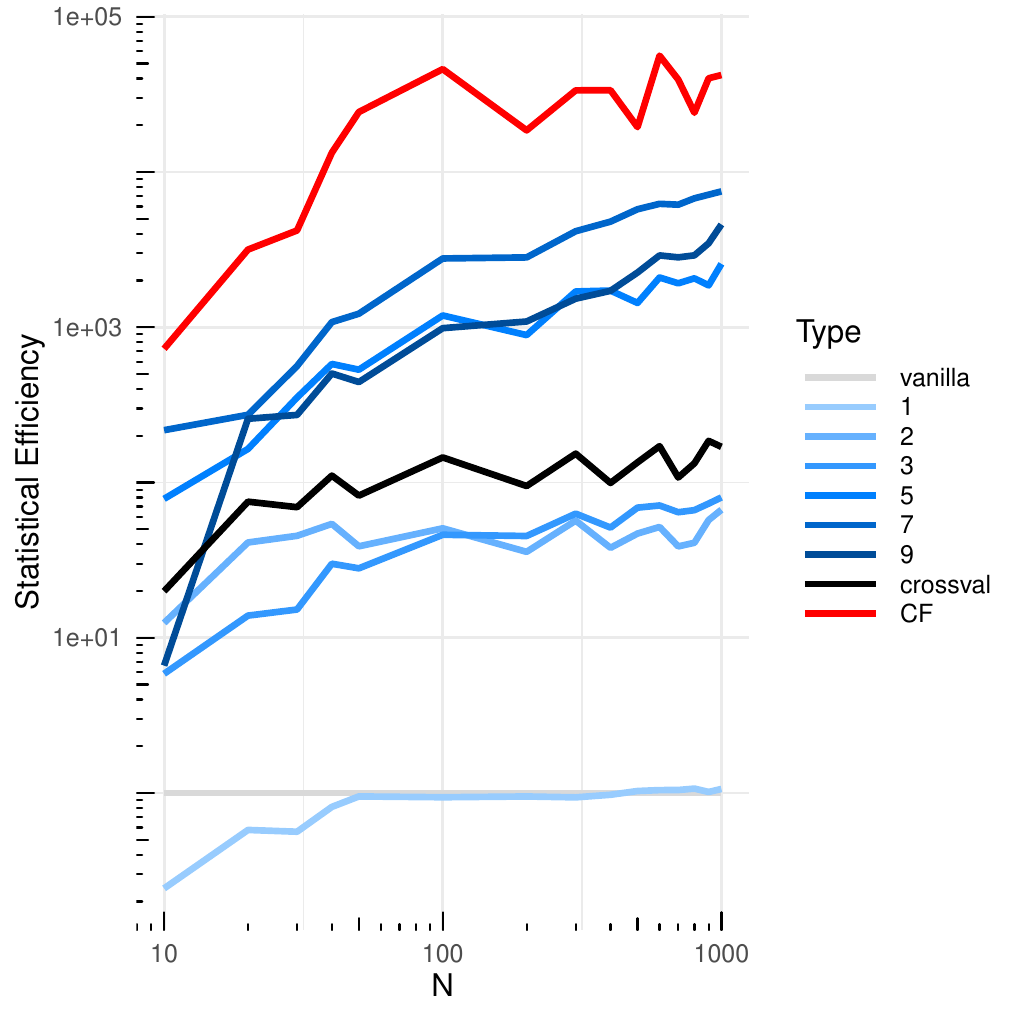}}
\subcaptionbox{Overall efficiency SMC\label{subfig:vdp_SMC_eff}}{\includegraphics[width=0.33\textwidth]{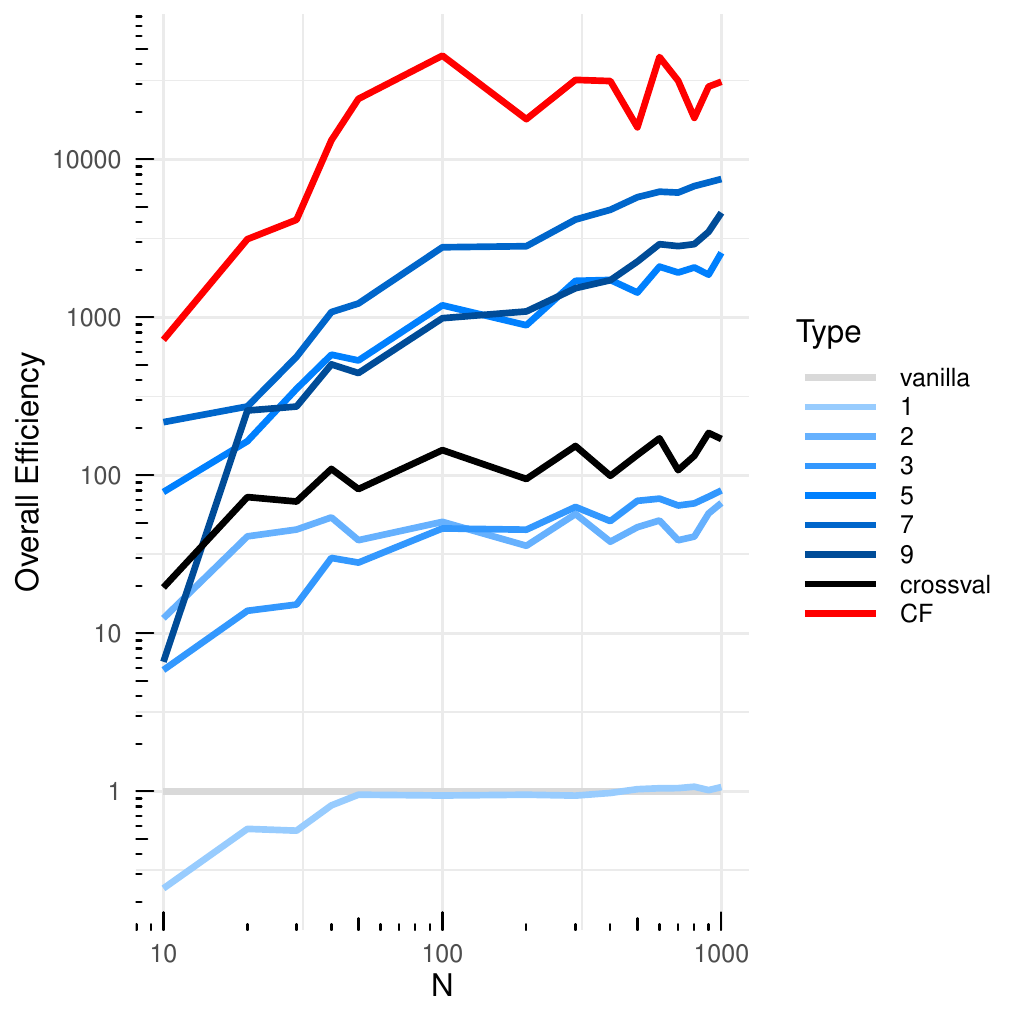}}
\caption{Van der Pol example: (a) statistical efficiency of the CTI estimator, (b) overall efficiency of the CTI estimator,  (c) statistical efficiency of the SMC estimator and (d) overall efficiency of the SMC estimator.}
\label{fig:VanDerPol_evidence}
\end{figure*}

Figure \ref{fig:VanDerPol_evidence} shows the statistical and overall efficiency of the two evidence estimators for different $N$. It is clear from these figures that higher order polynomials can offer substantial improvements in statistical efficiency over low order polynomials given a sufficient number of samples. CF performs similarly to the best performing ZV-CV method for the CTI estimator and typically slightly better than the best performing ZV-CV method for the SMC evidence estimator, at the cost of relatively poor performance for some values of $N$ due to outlying estimates. The overall trends are similar for overall efficiency, although the additional expectations lead to an increased cost in post\-processing which is especially noticeable for CF.

The cross-validation method results in estimators with similar MSE to fixed polynomial orders of $Q=2$ or $Q=3$. Although the chosen control variates lead to efficiency improvements of one to three orders of magnitude over the vanilla estimator, the improvements are not as substantial as for the optimal fixed $Q \in \{1,2,\ldots,9\}$. The cross-validation method is stopping at a sub-optimal $Q$, potentially because there is little to no improvement from one (sub-optimal) polynomial order to the next (sub-optimal) polynomial order for many of the expectations involved here. This hypothesis is supported by the polynomial order selection results in Table \ref{tab:VanDerPol_SMC_choice}. One potential solution in practice may be to force the cross-validation to compare up to at least some fixed polynomial order.

\begin{figure*}
\centering
\subcaptionbox{CTI $N=10$\label{subfig:vdp_CTI_10}}{\includegraphics[width=0.33\textwidth]{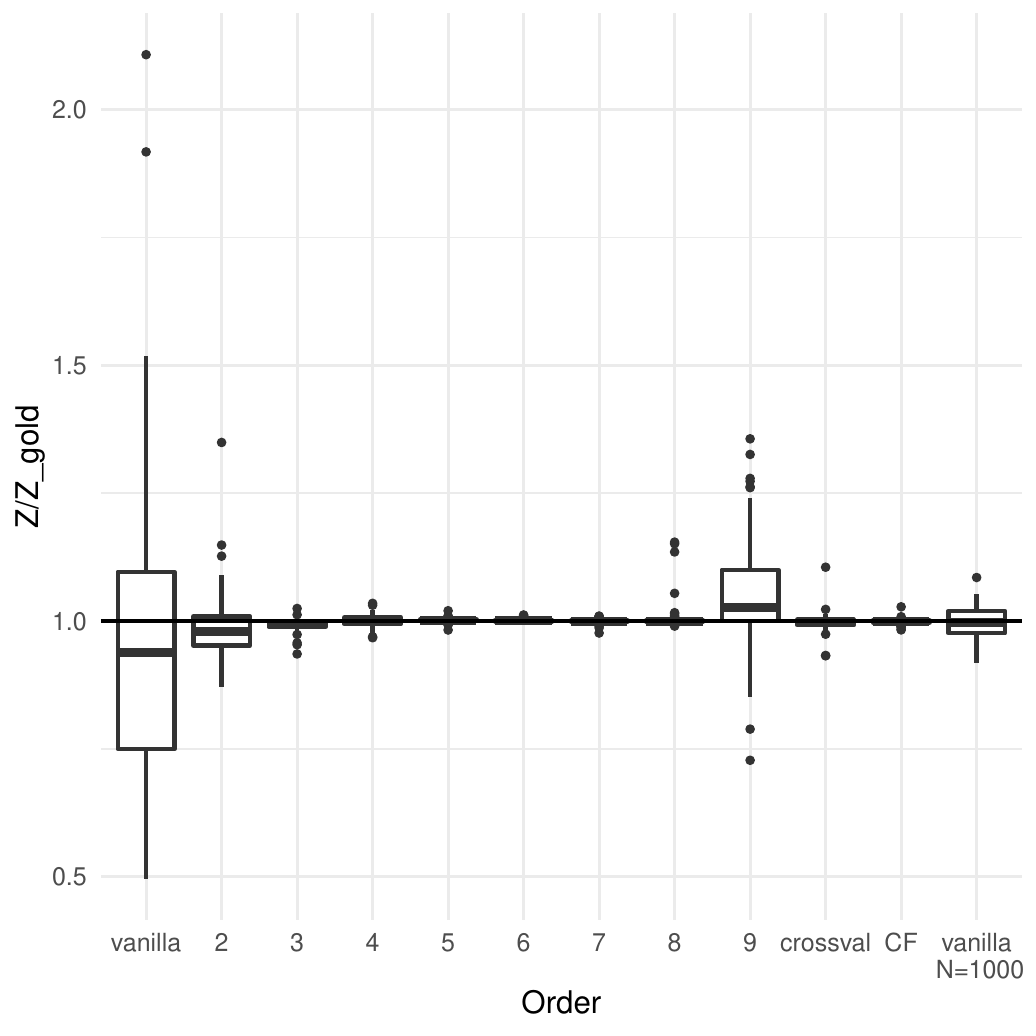}}
\subcaptionbox{SMC $N=10$\label{subfig:vdp_SMC_10}}{\includegraphics[width=0.33\textwidth]{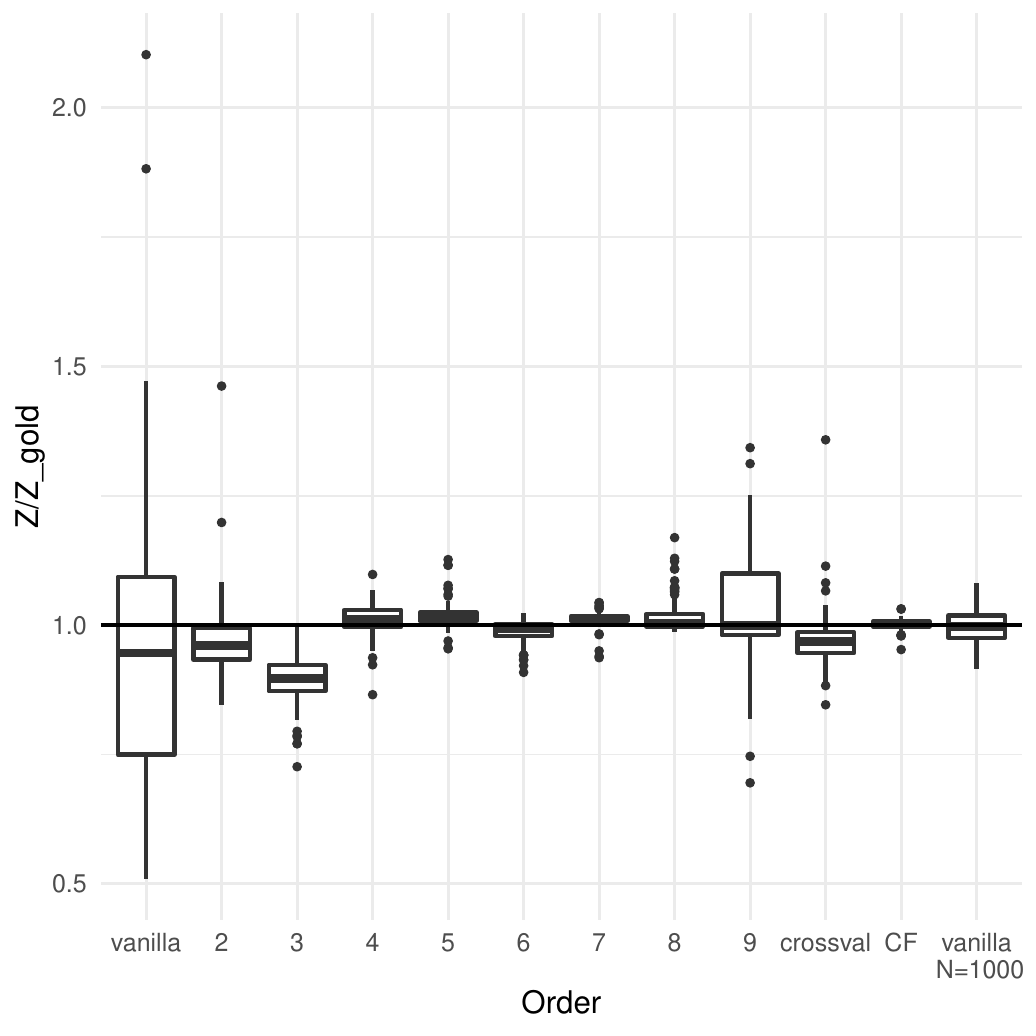}}
\caption{Van der Pol example: (a) performance of the CTI estimator for $N=10$ and (b) performance of the SMC estimator for $N=10$. The vanilla Monte Carlo estimator for $N=1000$ is shown for comparison.}
\label{fig:VanDerPol_evidence_10l}
\end{figure*}

At $N=10$, the evidence estimates for the CTI estimator appear to be approximately unbiased (Figure \ref{subfig:vdp_CTI_10}). The estimators also have low variance with the exception of the polynomial with $Q=9$, for which there are not enough samples for a reasonable fit. The SMC combined estimator suffers from more bias (Figure \ref{subfig:vdp_SMC_10}), but this bias disappears when the samples are split to facilitate independent estimation of $g(\vtheta)$ and evaluation of (2.1) of the main text (results not shown). 

\renewcommand{\tabcolsep}{3pt}

\begin{table}[!h]
\centering \footnotesize
\resizebox{\columnwidth}{!}{%
\begin{tabular}{rrrrrrrrrrrrrrrrrrrrrrrrrrr}
  \hline
  &  \multicolumn{26}{c}{SMC Iteration} \\
  \hline
$\mathbf{Q}$ & \textbf{1} & \textbf{2} & \textbf{3} & \textbf{4} & \textbf{5} & \textbf{6} & \textbf{7} & \textbf{8} & \textbf{9} & \textbf{10} & \textbf{11} & \textbf{12} & \textbf{13} & \textbf{14} & \textbf{15} & \textbf{16} & \textbf{17} & \textbf{18} & \textbf{19} & \textbf{20} & \textbf{21} & \textbf{22} & \textbf{23} & \textbf{24} & \textbf{25} & \textbf{26} \\ 
  \hline
\textbf{0} &   0 &   0 &   0 &   0 &   0 &   0 &   0 &   0 &   0 &   0 &   0 &   0 &   0 &   0 &   0 &   0 &   0 &   0 &   0 &   0 &   0 &   0 &   0 &   0 &   0 &   0 \\ 
\textbf{1} &   0 &   0 &   0 &   0 &   0 &   0 &   0 &   0 &   0 &   0 &   0 &   0 &   0 &   0 &   0 &   0 &   0 &   0 &   0 &   0 &   0 &   0 &   0 &   0 &   0 &   0 \\ 
\textbf{2} &  13 &  40 &  79 & 100 &   8 &  34 &  86 &  99 &  68 &  49 &  25 &  10 &  23 &  16 &  10 &   7 &  16 &  20 &  21 &  27 &  33 &  35 &  44 &  49 &  52 &   0 \\ 
\textbf{3} &   7 &   0 &   1 &   0 &   6 &   3 &   0 &   0 &   0 &   0 &   0 &   0 &   0 &   0 &   0 &   1 &   0 &   0 &   0 &   0 &   0 &   0 &   0 &   0 &   0 &   0 \\ 
\textbf{4}  &   7 &   0 &   0 &   0 &   0 &   0 &   0 &   0 &   1 &  14 &  23 &  34 &   5 &   7 &  10 &  17 &  12 &  17 &  21 &  11 &  21 &  22 &  27 &  18 &  16 &   0 \\ 
\textbf{5}  &  47 &  39 &  13 &   0 &  40 &  29 &  10 &   1 &  14 &  10 &  16 &  15 &  22 &  25 &  23 &  23 &  15 &  18 &  12 &   2 &   4 &   2 &   0 &   1 &   1 &   3 \\ 
\textbf{6}  &   0 &   1 &   0 &   0 &   0 &   0 &   0 &   0 &   0 &   0 &   0 &   0 &   0 &   0 &   0 &   0 &   0 &   0 &   0 &   0 &   0 &   0 &   0 &   1 &   1 &   0 \\ 
\textbf{7}  &   9 &  15 &   5 &   0 &  19 &  14 &   1 &   0 &   4 &   7 &   8 &  16 &  19 &  25 &  31 &  32 &  34 &  32 &  37 &  34 &  16 &  21 &  13 &   4 &   6 &  30 \\ 
\textbf{8}  &   5 &   0 &   0 &   0 &   5 &  10 &   1 &   0 &   8 &  11 &  16 &  15 &  11 &   7 &   5 &   2 &   0 &   0 &   0 &   0 &   0 &   0 &   0 &   0 &   0 &   0 \\ 
\textbf{9} &   0 &   1 &   0 &   0 &   0 &   0 &   0 &   0 &   0 &   0 &   0 &   0 &   0 &   0 &   0 &   0 &   0 &   2 &   2 &  13 &  15 &  11 &  11 &  18 &  16 &   0 \\ 
\textbf{10}  &   3 &   2 &   2 &   0 &   4 &   0 &   0 &   0 &   2 &   2 &   3 &   2 &  12 &  14 &  10 &  12 &  16 &   7 &   4 &   1 &   0 &   0 &   0 &   0 &   0 &  55 \\ 
\textbf{11}  &   0 &   0 &   0 &   0 &   9 &   7 &   2 &   0 &   2 &   1 &   5 &   3 &   2 &   0 &   3 &   0 &   0 &   0 &   0 &   0 &   0 &   0 &   0 &   3 &   3 &   0 \\ 
\textbf{12}  &   4 &   1 &   0 &   0 &   1 &   0 &   0 &   0 &   0 &   0 &   0 &   0 &   0 &   2 &   0 &   0 &   5 &   3 &   1 &   7 &   6 &   3 &   1 &   1 &   1 &   1 \\ 
\textbf{13} &   2 &   0 &   0 &   0 &   0 &   0 &   0 &   0 &   0 &   2 &   2 &   1 &   5 &   2 &   4 &   4 &   0 &   0 &   0 &   0 &   0 &   0 &   0 &   0 &   1 &   0 \\ 
\textbf{14} &   3 &   1 &   0 &   0 &   4 &   2 &   0 &   0 &   0 &   0 &   0 &   1 &   0 &   0 &   1 &   0 &   0 &   0 &   0 &   0 &   0 &   2 &   2 &   1 &   2 &   0 \\ 
\textbf{15} &   0 &   0 &   0 &   0 &   1 &   0 &   0 &   0 &   0 &   0 &   0 &   0 &   0 &   0 &   2 &   1 &   1 &   1 &   0 &   0 &   0 &   0 &   0 &   0 &   0 &   8 \\ 
\textbf{16} &   0 &   0 &   0 &   0 &   0 &   0 &   0 &   0 &   0 &   2 &   2 &   1 &   0 &   1 &   0 &   0 &   0 &   0 &   0 &   0 &   1 &   0 &   0 &   1 &   0 &   3 \\ 
\textbf{17} &   0 &   0 &   0 &   0 &   2 &   0 &   0 &   0 &   0 &   0 &   0 &   1 &   0 &   0 &   0 &   0 &   0 &   0 &   1 &   4 &   3 &   1 &   2 &   1 &   0 &   0 \\ 
\textbf{18} &   0 &   0 &   0 &   0 &   0 &   0 &   0 &   0 &   1 &   1 &   0 &   1 &   0 &   1 &   1 &   1 &   1 &   0 &   0 &   0 &   0 &   0 &   0 &   1 &   0 &   0 \\ 
\textbf{19} &   0 &   0 &   0 &   0 &   0 &   1 &   0 &   0 &   0 &   1 &   0 &   0 &   0 &   0 &   0 &   0 &   0 &   0 &   0 &   1 &   1 &   1 &   0 &   1 &   0 &   0 \\ 
\textbf{20} &   0 &   0 &   0 &   0 &   1 &   0 &   0 &   0 &   0 &   0 &   0 &   0 &   1 &   0 &   0 &   0 &   0 &   0 &   1 &   0 &   0 &   1 &   0 &   0 &   1 &   0 \\ 
\textbf{21} &   0 &   0 &   0 &   0 &   0 &   0 &   0 &   0 &   0 &   0 &   0 &   0 &   0 &   0 &   0 &   0 &   0 &   0 &   0 &   0 &   0 &   1 &   0 &   0 &   0 &   0 \\ 
   \hline
\end{tabular}}
\caption{Van der Pol example: The chosen polynomial order counts across 100 independent runs using cross-validation for each of the 26 expectations in the SMC evidence estimator. Results are based on $N=1000$.}
\label{tab:VanDerPol_SMC_choice}
\end{table}

\normalsize
\clearpage

\section{Ordinary Differential Equation Example}\label{app:ODE}
Potential limitations of ZV-CV and regularized ZV-CV are illustrated through this challenging example with non-linear posterior dependencies. For the mostpart, the performance with low sample sizes is poor and the performance with larger sample sizes is modest compared to the examples in Section 4.1 of the main paper and Appendices D and E of these Online Resources. The likelihood function is relatively expensive in this example, but we consider statistical efficiency only because this is sufficient to illustrate the potential limitations of ZV-CV and regularized ZV-CV.

\citet{Geyer1991a} describes the following system of coupled ordinary differential equations (ODEs) for modelling biochemical pathways,
\begin{align*}
\begin{matrix}
\frac{dS}{dt}=-k_1 S \\[0.5em]
\frac{dD}{dt}=k_1 S \\[0.5em]
\frac{dR}{dt}=-\frac{V_1 R S}{K_{m_1}+R}+\frac{V_2 R_{pp}}{K_{m_2}+R_{pp}} \\[0.5em]
\frac{dR_{pp}}{dt}=\frac{V_1 R S}{K_{m_1}+R}-\frac{V_2 R_{pp}}{K_{m_2}+R_{pp}}.
\end{matrix}
\end{align*}

This example has been considered from a Bayesian context in \citet{Girolami2008}, \citet{South2018} and \citet{Salomone2018}. Following \citet{South2018}, observations $y(t)$ of $R_{pp}(t)$ are observed with noise such that $y(t) \sim \mathcal{N}(R_{pp}(t),0.02^2)$ for $t=0,3,\ldots,57$. The nine parameters of interest are $\vtheta = (k_1, V_1, K_{m_1}, K_{m_2}, V_2, S_0, D_0, R_0, R_{pp_0})$, the last four being the initial values of $S$, $D$, $R$ and $R_{pp}$. We use the same simulated data as \citet{South2018} which is based on $\vtheta=(0.05,0.20,0.1,0.1,0.1,1.00,0,1.00,0)$ and we use the priors from \citet{Girolami2008},

\begin{center}
$k_1$, $V_1$, $K_{m_1}$, $V_2$, $K_{m_2} \sim \mathcal{G}(1,1)$

$S_0$, $R_0 \sim \mathcal{G}(5,0.2)$

$D_0$, $R_{pp_0} \sim \mathcal{G}(1,0.1)$,
\end{center}
where $\mathcal{G}(a,b)$ represents the Gamma distribution with shape parameter $a$, scale parameter $b$ and mean $ab$. The log-transformed parameters, $\vect{\psi} = \log \vtheta$, are used for MCMC proposals and for ZV-CV through (2.5) of the main paper.

As explained in \citet{South2018}, this example has non-linear posterior dependencies (see Appendix G of \citet{South2018} for an illustrative figure) and is challenging partially because of numerous practical and structural identifiability issues in the model. It is also interesting to note that no information is obtained about $D_0$ by observing $R_{pp}$, so $D_0$ is practically non-identifiable. $D_0$ is therefore independent of other parameters and of the data, so the posterior marginal for $D_0$ is simply its prior, $\mathcal{G}(1,0.1)$. This helps to explain the extremely good performance of ZV-CV and subset regularized ZV-CV in estimating its posterior mean. The parameter $R_{pp_0}$ is well identified based on the data and is therefore conditionally independent of other parameters given the data, making it another parameter which is potentially easier to estimate.

The gold standard of posterior expectation and evidence estimation is the mean from 100 fixed SMC runs with $N=10000$, except for the posterior mean for $D_0$ for which we use the true value of $0.1$. When using the SMC evidence estimator, it is possible to obtain a negative estimate for the evidence and this happened in some runs with low $N$ for this example. To avoid this issue, we repeat the coefficient estimation with a fixed intercept of $\hat{c}=\frac{1}{N}\sum_{i=1}^N \varphi(\vtheta_i)$ when the evidence estimate is negative. For CF, we instead replace the estimator with $\frac{1}{N}\sum_{i=1}^N \varphi(\vtheta_i)$ in this case, which amounts to using vanilla MC.

\subsubsection{Posterior Expectations}
Tables \ref{tab:ODE1}-\ref{tab:ODE9} show the statistical efficiency for each posterior expectation. Given the complex target distribution and strong dependencies between parameters, it is not possible to achieve improvements on vanilla Monte Carlo integration using $N<100$ for most parameters. CF outperforms ZV-CV for some marginals, but the improvements are mostly similar to ZV-CV. The worst performing marginal expectation improves on vanilla Monte Carlo integration by at factor of at most 5.4 at $N=1000$.

There is generally no difference in efficiency between vanilla Monte Carlo integration and \textit{a priori} ZV-CV, with the exceptions of $D_0$ and $R_{pp_0}$. Even for a sample size as small as $N=10$, using ZV-CV either with a full polynomial in $\vect{\psi}$ or with a polynomial only in $\log D_0$ gives a posterior mean for $D_0$ that is correct to 15 significant figures. The improvement is 30 orders of magnitude smaller for the same expectation with CF. Using a polynomial in $\log R_{pp_0}$ gives an estimate of $\overline{R_{pp_0}}$ which is 1.5 times more efficient than vanilla Monte Carlo integration at $N=1000$.

\begin{table}[!h]
\centering \footnotesize
\resizebox{\columnwidth}{!}{%
\begin{tabular}{rrrrrrrrrrrrrrrr}
  \hline
$N$ & \zv{1} & \zvl{1} & \zvr{1} & \zv{2} & \zvl{2} & \zvr{2} & \zv{3} & \zvl{3} & \zvr{3} & \zv{4} & \zvl{4} & \zvr{4} & \zvs{1}{1} & crossval  & CF\\ 
  \hline
50 & 0.55 & 0.63 & 0.62 &  & 0.93 & 0.93 &  & 0.72 & 0.79 &  & 0.70 & 0.77 & 1.0 & 0.76 & \textbf{2.0} \\ 
  100 & 0.80 & 0.85 & 0.84 & 2.0 & \textbf{2.3} & 1.7 &  & 0.92 & 0.76 &  & 0.94 & 0.86 & 1.0 & 1.7  & 1.8 \\ 
  500 & 0.97 & 0.97 & 0.97 & \textbf{8.7} & 5.3 & 6.0 & 3.6 & 7.6 & 5.3 &  & 7.2 & 4.3 & 1.0 & 6.2  & 6.9\\ 
  1000 & 0.92 & 0.95 & 0.94 & 5.7 & 5.0 & \textbf{5.9} & 4.5 & 5.3 & 4.4 & 0.97 & 4.5 & 3.5 & 1.0 & 4.3 & 5.0 \\ 
   \hline
\end{tabular}}
\caption{ODE example, marginal 1}
\label{tab:ODE1}
\end{table}

\begin{table}[!h]
\centering \footnotesize
\resizebox{\columnwidth}{!}{%
\begin{tabular}{rrrrrrrrrrrrrrrr}
  \hline
$N$ & \zv{1} & \zvl{1} & \zvr{1} & \zv{2} & \zvl{2} & \zvr{2} & \zv{3} & \zvl{3} & \zvr{3} & \zv{4} & \zvl{4} & \zvr{4} & \zvs{1}{1} & crossval  & CF \\ 
  \hline
50 & 0.56 & 0.98 & 1.2 &  & 0.34 & 0.34 &  & 0.66 & 0.70 &  & 0.78 & 0.79 & \textbf{1.0} & 0.55  & 0.58 \\ 
  100 & 1.7 & 1.9 & 2.0 & 2.0 & 0.80 & 0.86 &  & 0.90 & 0.77 &  & 1.0 & 0.92 & 1.0 & 1.9  & \textbf{3.2}\\ 
  500 & 1.7 & 1.7 & 1.7 & 4.1 & 1.4 & 2.9 & 7.4 & 2.3 & 3.3 &  & 1.7 & 1.7 & 1.0 & 7.2  & \textbf{8.2} \\ 
  1000 & 1.1 & 1.4 & 1.4 & 1.8 & 1.1 & 1.9 & 3.8 & 1.4 & 2.4 & 1.7 & 1.6 & 1.6 & 1.0 & 3.8 & \textbf{5.4}  \\ 
   \hline
\end{tabular}}
\caption{ODE example, marginal 2} 
\end{table}

\begin{table}[!h]
\centering \footnotesize
\resizebox{\columnwidth}{!}{%
\begin{tabular}{rrrrrrrrrrrrrrrr}
  \hline
$N$ & \zv{1} & \zvl{1} & \zvr{1} & \zv{2} & \zvl{2} & \zvr{2} & \zv{3} & \zvl{3} & \zvr{3} & \zv{4} & \zvl{4} & \zvr{4} & \zvs{1}{1} & crossval  & CF\\ 
  \hline
50 & 0.60 & 0.70 & 0.75 &  & 0.30 & 0.35 &  & 0.90 & 0.68 &  & 1.2 & 0.44 & 1.0  & 0.29 & \textbf{2.3} \\ 
  100 & 1.1 & 1.1 & 1.1 & 1.1 & 1.5 & 1.3 &  & 1.3 & 1.1 &  & 1.4 & 1.2 & 1.0 & 1.2  & \textbf{2.2} \\ 
  500 & 1.4 & 1.3 & 1.3 & 4.6 & 1.7 & 2.4 & \textbf{19} & 3.7 & 4.0 &  & 2.7 & 2.2 & 1.0 & 5.6 & 17  \\ 
  1000 & 1.3 & 1.2 & 1.3 & 1.2 & 1.3 & 1.3 & 5.1 & 2.1 & 2.3 & 2.7 & 2.1 & 1.5 & 1.0 & 4.3  & \textbf{8.8}\\ 
   \hline
\end{tabular}}
\caption{ODE example, marginal 3} 
\end{table}

\begin{table}[!h]
\centering \footnotesize
\resizebox{\columnwidth}{!}{%
\begin{tabular}{rrrrrrrrrrrrrrrr}
  \hline
$N$ & \zv{1} & \zvl{1} & \zvr{1} & \zv{2} & \zvl{2} & \zvr{2} & \zv{3} & \zvl{3} & \zvr{3} & \zv{4} & \zvl{4} & \zvr{4} & \zvs{1}{1} & crossval  & CF\\ 
  \hline
50 & 0.51 & 0.60 & 0.58 &  & 0.35 & 0.34 &  & 0.37 & 0.41 &  & 0.34 & 0.51 & 1.0 & 0.39  & \textbf{2.4} \\ 
  100 & 0.91 & 0.89 & 0.88 & 1.7 & 1.5 & 0.99 &  & \textbf{1.9} & 1.3 &  & 1.3 & 1.2 & 1.0 & 1.7  & 0.90\\ 
  500 & 1.2 & 1.2 & 1.2 & 5.1 & 4.8 & 4.7 & 3.0 & 7.8 & 7.0 &  & 5.9 & 5.7 & 1.0 & 5.7  & \textbf{8.0}\\ 
  1000 & 1.2 & 1.2 & 1.2 & 5.4 & 5.4 & 6.0 & 4.2 & \textbf{11} & 11 & 0.41 & 9.0 & 7.8 & 1.0 & 5.5 & 6.2 \\ 
   \hline
\end{tabular}}
\caption{ODE example, marginal 4} 
\end{table}

\begin{table}[!h]
\centering \footnotesize
\resizebox{\columnwidth}{!}{%
\begin{tabular}{rrrrrrrrrrrrrrrr}
  \hline
$N$ & \zv{1} & \zvl{1} & \zvr{1} & \zv{2} & \zvl{2} & \zvr{2} & \zv{3} & \zvl{3} & \zvr{3} & \zv{4} & \zvl{4} & \zvr{4} & \zvs{1}{1} & crossval & CF \\ 
  \hline
50 & 0.62 & 0.69 & 0.69 &  & 0.96 & 0.98 &  & 0.85 & 0.86 &  & 0.86 & 0.88 & 1.0 & 0.80  & \textbf{1.6} \\ 
  100 & 0.85 & 0.86 & 0.85 & 1.8 & \textbf{1.8} & 1.7 &  & 0.88 & 1.0 &  & 0.86 & 0.90 & 1.0 & 1.4 &  1.6\\ 
  500 & 0.99 & 0.99 & 0.99 & 6.9 & 3.8 & 6.2 & 6.1 & \textbf{9.9} & 7.9 &  & 6.7 & 7.4 & 1.0 & 7.3 &  5.9\\ 
  1000 & 0.98 & 1.0 & 0.99 & 6.6 & 5.4 & 7.0 & 7.3 & 9.9 & 8.9 & 2.0 & \textbf{11} & 9.1 & 1.0 & 8.0 &  7.1\\ 
   \hline
\end{tabular}}
\caption{ODE example, marginal 5} 
\end{table}

\begin{table}[!h]
\centering \footnotesize
\resizebox{\columnwidth}{!}{%
\begin{tabular}{rrrrrrrrrrrrrrrr}
  \hline
$N$ & \zv{1} & \zvl{1} & \zvr{1} & \zv{2} & \zvl{2} & \zvr{2} & \zv{3} & \zvl{3} & \zvr{3} & \zv{4} & \zvl{4} & \zvr{4} & \zvs{1}{1} & crossval  & CF\\ 
  \hline
50 & 39 & 1.8 & 2.1 &  & 0.62 & 0.70 &  & 0.71 & 0.53 &  & 0.84 & 0.91 & 1.0 & \textbf{39}  & 31 \\ 
  100 & 55 & 5.3 & 7.9 & 59 & 1.7 & 1.9 &  & 0.85 & 0.73 &  & 0.88 & 0.87 & 1.0 & 64 &  \textbf{78}\\ 
  500 & 63 & 8.9 & 14 & \textbf{69} & 5.3 & 23 & 55 & 4.6 & 9.9 &  & 3.5 & 2.6 & 1.0 & 54  & 51 \\ 
  1000 & \textbf{52} & 3.9 & 6.4 & 44 & 2.9 & 19 & 32 & 3.7 & 8.7 & 35 & 3.4 & 3.0 & 1.0 & 32 & 29 \\ 
   \hline
\end{tabular}}
\caption{ODE example, marginal 6} 
\end{table}

\begin{table}[!h]
\centering \footnotesize
\resizebox{\columnwidth}{!}{%
\begin{tabular}{rrrrrrrrrrrr}
  \hline
$N$ & \zv{1} & \zvl{1} & \zv{2} & \zvl{2} & \zv{3} & \zvl{3} & \zv{4} & \zvl{4} & \zvs{1}{1} & crossval & CF \\  
  \hline
50 & 80 & 97 &  & 120 &  & 78 &  & \textbf{860} & 210 &  98  & 2.5 $\times 10^{-28}$\\ 
  100 & 6.3 & 6.2 & 4.2 & 6.1 &  & 6.7 &  & 7.2 & \textbf{7.3} & 7.1 &  7.4 $\times 10^{-25}$ \\ 
  500 & 15 & 5.9 & 17 & 13 & 3.0 & 13 &  & \textbf{18} & 17 & 16 & 1.6 $\times 10^{-23}$  \\ 
  1000 & 14 & 5.5 & \textbf{15} & 9.2 & 8.6 & 11 & 1.0 & 14 & 14 & 14 &  4.8 $\times 10^{-23}$\\ 
   \hline
\end{tabular}}
\caption{ODE example, marginal 7. Results are scaled down by a factor of $10^{-29}$. That is, the smallest and largest improvements are $25$ and $8.6 \times 10^{31}$, respectively. Ridge regression is excluded from this table because its performance is far worse than LASSO (by a factor of at least $10^{22}$ for all $Q$ and $N$).} 
\end{table}

\begin{table}[!h]
\centering \footnotesize
\resizebox{\columnwidth}{!}{%
\begin{tabular}{rrrrrrrrrrrrrrrr}
  \hline
$N$ & \zv{1} & \zvl{1} & \zvr{1} & \zv{2} & \zvl{2} & \zvr{2} & \zv{3} & \zvl{3} & \zvr{3} & \zv{4} & \zvl{4} & \zvr{4} & \zvs{1}{1} & crossval & CF \\ 
  \hline
50 & 0.91 & 0.90 & 0.86 &  & 0.69 & 0.85 &  & 1.1 & 1.1 &  & 1.0 & \textbf{1.2} & 0.99 & 0.74  & 1.1\\ 
  100 & 1.3 & 1.3 & 1.3 & 1.6 & \textbf{1.8} & 1.6 &  & 1.5 & 1.5 &  & 1.1 & 1.1 & 1.0 & 1.8  & 0.90\\ 
  500 & 1.5 & 1.5 & 1.5 & 2.5 & 2.5 & 2.3 & 2.4 & 5.0 & \textbf{5.1} &  & 3.0 & 3.1 & 1.0 & 3.4 & 3.3 \\ 
  1000 & 1.5 & 1.5 & 1.5 & 2.3 & 2.4 & 2.3 & 2.0 & \textbf{6.7} & 6.6 & 0.76 & 4.5 & 4.3 & 1.0 & 2.4 & 1.9 \\ 
   \hline
\end{tabular}}
\caption{ODE example, marginal 8} 
\end{table}

\begin{table}[!h]
\centering 
\resizebox{\columnwidth}{!}{%
\begin{tabular}{rrrrrrrrrrrrrrrr}
  \hline
$N$ & \zv{1} & \zvl{1} & \zvr{1} & \zv{2} & \zvl{2} & \zvr{2} & \zv{3} & \zvl{3} & \zvr{3} & \zv{4} & \zvl{4} & \zvr{4} & \zvs{1}{1} & crossval & CF \\ 
  \hline
50 & 2.3 & 2.4 & 2.3 &  & 5.1 & 6.4 &  & \textbf{7.8} & 5.9 &  & 5.7 & 4.0 & 1.3 & 4.6  & 3.0\\ 
  100 & 2.0 & 2.0 & 2.0 & 5.3 & 5.7 & 5.5 &  & \textbf{7.4} & 6.0 &  & 5.6 & 5.4 & 1.3 & 5.7 & 5.2 \\ 
  500 & 2.5 & 2.4 & 2.4 & 29 & 32 & 30 & 50 & 48 & 57 &  & 31 & 41 & 1.4 & 48  & \textbf{75}\\ 
  1000 & 3.4 & 3.2 & 3.2 & 47 & 66 & 60 & 54 & 67 & 68 & 5.7 & 36 & 37 & 1.5 & 55  & \textbf{96}\\ 
   \hline
\end{tabular}}
\caption{ODE example, marginal 9}
\label{tab:ODE9} 
\end{table}

\subsubsection{Evidence Estimation}

\begin{figure}[!h]
\centering
\subcaptionbox{Vanilla SMC evidence estimates\label{subfig:ODE_SS_vanilla}}{\includegraphics[width=0.4\textwidth]{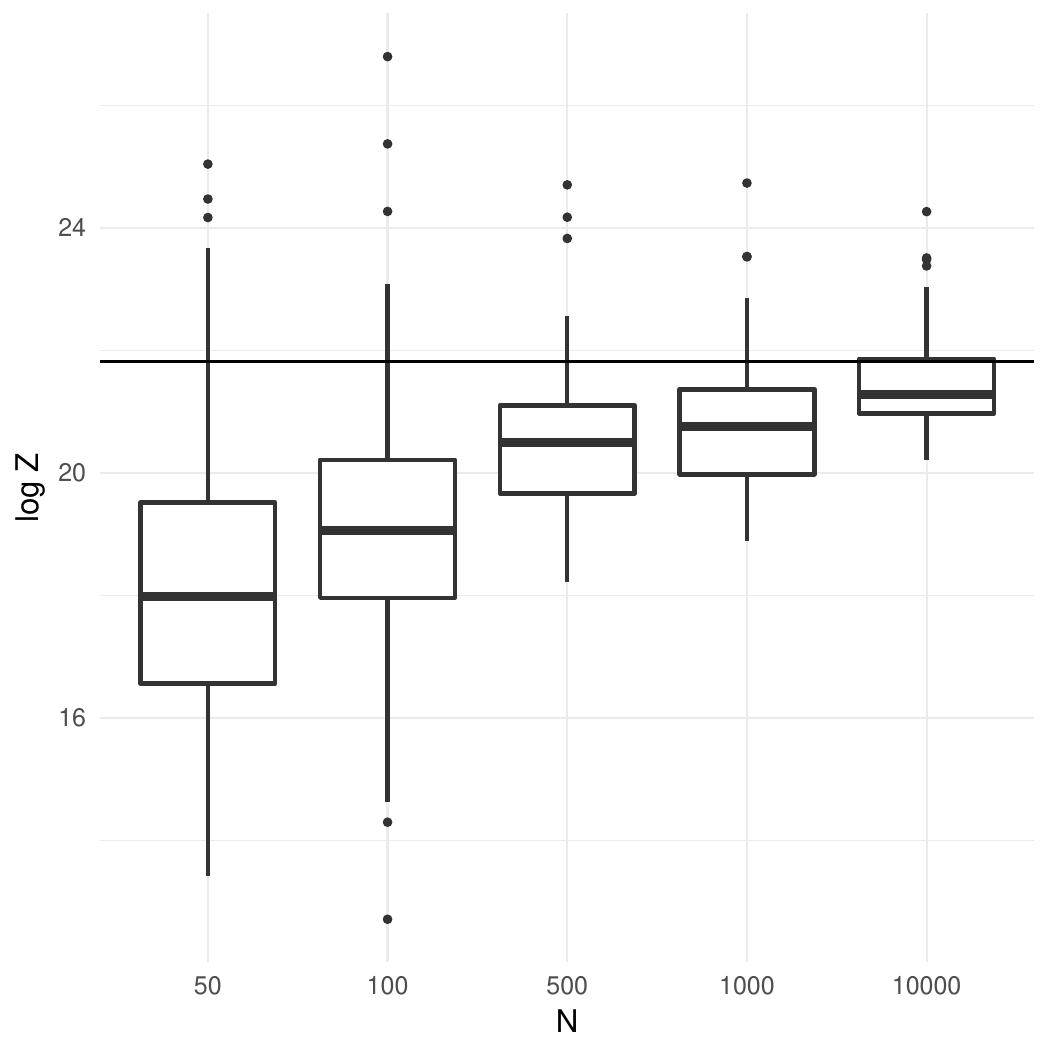} }
\subcaptionbox{SMC evidence estimates for $N=1000$\label{subfig:ODE_SS_bias}}{\includegraphics[width=0.4\textwidth]{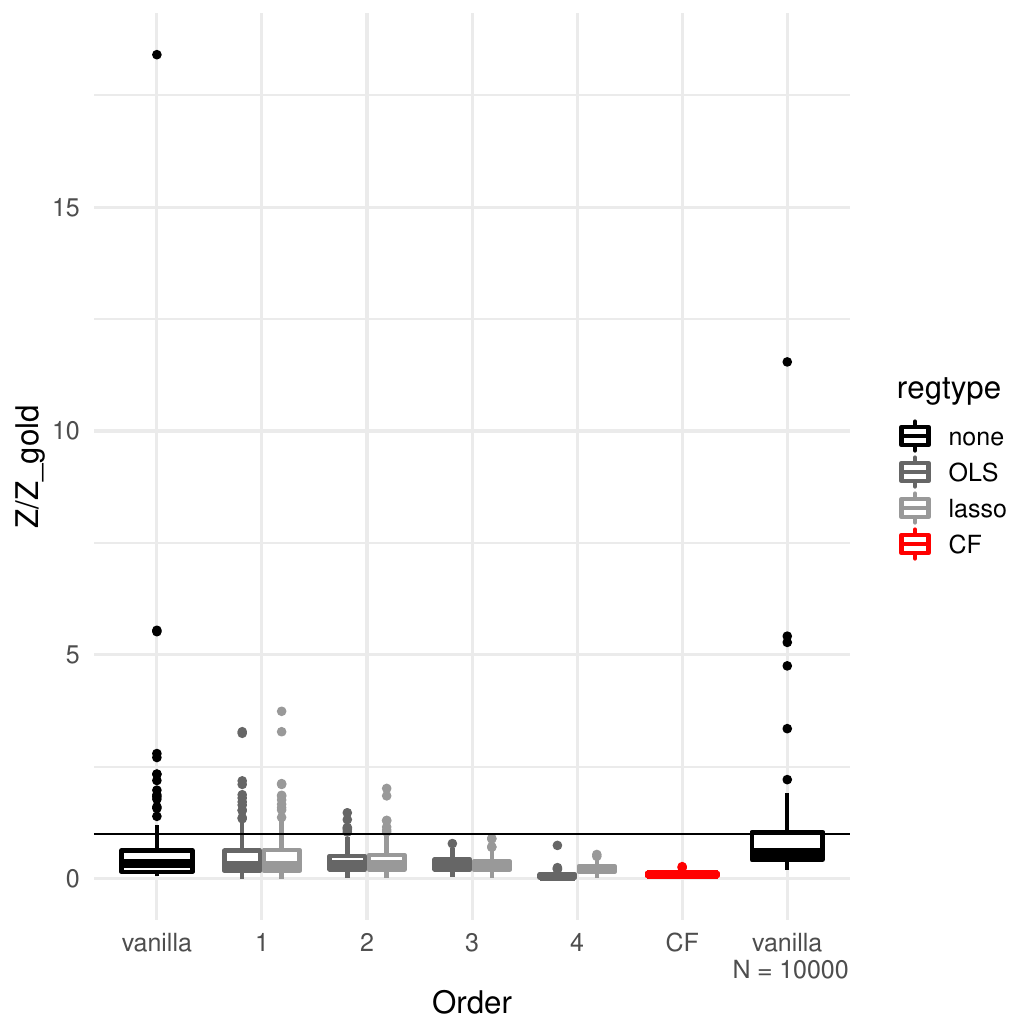} }
\caption{ODE example: (a) Vanilla Monte Carlo SMC evidence estimates for increasing $N$ and (b) SMC evidence estimates for $N=1000$, with results for $N=10000$ vanilla Monte Carlo shown for comparison.}
\label{fig:ODE_evidence_issues}
\end{figure}

The vanilla Monte Carlo estimators of the evidence have very high variance in this example. From Figure \ref{fig:ODE_evidence_issues}, it can be seen that the distribution of SMC evidence estimates using vanilla Monte Carlo is positively skewed. The estimator underestimates with a high probability and overestimates by a large amount with small probability. The behaviour of the CTI estimator is similar.

Figure \ref{subfig:ODE_SS_bias} illustrates the performance of ZV-CV and CF for $N=1000$ compared to vanilla Monte Carlo integration with the gold standard of $N=10000$. ZV-CV and CF significantly reduce the variance, but they introduce a negative bias. As suspected, the MSE is reduced by using ZV-CV and CF but the amount by which it is reduced is not stable across $N$ (Tables \ref{fig:ODE_CTI} and \ref{fig:ODE_SMC}) owing to the introduction of bias. Further research may be required to investigate the performance of ZV-CV and CF when the vanilla estimator is highly positively skewed.

Despite the challenges in this example, automatic control variate selection performs reasonably well.

\begin{table}[!h]
\centering
\caption{ODE example: statistical efficiency of the CTI evidence estimates for a range of $N$. Ridge regression performs either similarly or worse than LASSO and has therefore been excluded from this table. The * indicates that cross-validation efficiency values for $N=100$, $N=500$ and $N=1000$ are based on 98, 99 and 97 runs, respectively. The remaining 6 runs reached a polynomial order which required too much RAM.} 
\label{fig:ODE_CTI}
\begin{tabular}{rrrrrrrrrrr}
  \hline
$N$ & \zv{1} & \zvl{1} & \zv{2} & \zvl{2} & \zv{3} & \zvl{3} & \zv{4} & \zvl{4} & crossval & CF \\ 
  \hline
  50 & 29 & 2.2 &  & 33 &  & 30 &  & 59 & 29 & \textbf{100}\\ 
  100 & 28 & 5.0 & 210 & 50 &  & 3.5 &  & 1.7 & \textbf{240*} & 220\\ 
  500 & 6.0 & 5.7 & 8.4 & \textbf{9.0} & 7.1 & 7.3 &  & 6.6 & 8.1* & 6.3 \\ 
  1000 & 7.3 & 7.0 & \textbf{10} & 9.9 & 9.3 & 8.7 & 4.9 & 7.3 & 9.6* & 5.5 \\ 
   \hline
\end{tabular}
\end{table}

\begin{table}[!h]
\centering
\caption{ODE example: statistical efficiency of the SMC evidence estimates for a range of $N$. Ridge regression performs either similarly or worse than LASSO and has therefore been excluded from this table.} 
\label{fig:ODE_SMC}
\begin{tabular}{rrrrrrrrrrr}
  \hline
$N$ & \zv{1} & \zvl{1} & \zv{2} & \zvl{2} & \zv{3} & \zvl{3} & \zv{4} & \zvl{4} & crossval & CF \\ 
  \hline
  50 & 5.6 & 3.4 &  & 9.2 &  & 9.3 &  & 9.1 & 5.4  & \textbf{9.8}\\ 
  100 & 33 & 7.1 & \textbf{240} & 28 &  & 1.7 &  & 3.0 & 29 & 220 \\ 
  500 & 5.5 & 5.3 & 7.8 & \textbf{8.2} & 6.9 & 7.1 &  & 6.2 & 7.3 & 5.3 \\ 
  1000 & 6.9 & 6.6 & \textbf{9.1} & 9.0 & 8.6 & 8.1 & 4.7 & 6.8 & 8.1& 5.0 \\ 
   \hline
\end{tabular}
\end{table}

\end{document}